\documentclass[12pt]{iopart}

\usepackage{graphicx}
\usepackage{tcolorbox}
\usepackage{float}
\usepackage[style=authoryear, maxnames=2, backend=biber, hyperref=true]{biblatex}

\usepackage{hyperref}
\hypersetup{
    colorlinks=true,
    linkcolor=blue,
    citecolor=blue,
    filecolor=magenta,
    urlcolor=cyan,
}
\usepackage[labelfont=bf]{caption}
\usepackage{subcaption}
\usepackage{enumitem}
\usepackage{comment}
\usepackage{multirow}

\begin{document}

\title[TTV]{The geography of inequalities in access to healthcare across England: the role of bus travel time variability}

\author{Zihao Chen$^1$, Federico Botta$^{2,3}$}

\address{$^1$ Department of Mathematics and Statistics, University of Exeter, North Park Road, Exeter, UK. EX4 4QF.}
\address{$^2$ Department of Computer Science, University of Exeter, North Park Road, Exeter, UK. EX4 4QF.}
\address{$^3$ The Alan Turing Institute, UK}
\ead{zihao.chen@exeter.ac.uk}
\vspace{10pt}

\begin{abstract}
Fair access to healthcare facilities is fundamental to achieving social equity. Traditional travel time-based accessibility measures often overlook the dynamic nature of travel times resulting from different departure times, which compromises the accuracy of these measures in reflecting the true accessibility experienced by individuals. This study examines public transport-based accessibility to healthcare facilities across England from the perspective of travel time variability (TTV). Using comprehensive bus timetable data from the Bus Open Data Service (BODS), we calculated hourly travel times from each Lower Layer Super Output Area (LSOA) to the nearest hospitals and general practices and developed a TTV metric for each LSOA and analysed its geographical inequalities across various spatial scales. Our analysis reveals notable spatial-temporal patterns of TTV and average travel times, including an urban-rural divide, clustering of high and low TTV regions, and distinct outliers. Furthermore, we explored the relationship between TTV and deprivation, categorising LSOAs into four groups based on their unique characteristics, which provides valuable insights for designing targeted interventions. Our study also highlights the limitations of using theoretical TTV derived from timetable data and emphasises the potential of using real-time operational data to capture more realistic accessibility measures. By offering a more dynamic perspective on accessibility, our findings complement existing travel time-based metrics and pave way for future research on TTV-based accessibility using real-time data. This evidence-based approach can inform efforts to ``level up" public transport services, addressing geographical inequalities and promoting equitable access to essential healthcare services.

\end{abstract}

%
\vspace{2pc}

\section{Introduction}
In human geography and transport research, accessibility refers to the ease with which people can reach desired destinations or opportunities (e.g., employment, education, healthcare, etc.) using a particular transport system (\cite{dalvi_measurement_1976}). This `ease' is comprised of several aspects, including the amount of time (travel, waiting, parking), cost (fixed, variable), and effort (convenience, reliabilty, comfort, etc.) (\cite{geurs_accessibility_2004}). Despite accessibility being multidimensional, travel time is often treated as synonymous with it and frequently used as a proxy in accessibility measurements for its good interpretability (\cite{nelson_a_travel_2008,weiss_global_2018}). Travel time to destinations is often the most critical determinant of an individual's ability to access opportunities, thus influencing people's modal choices (\cite{noland_simulated_2000,lucas_transport_2012}). Furthermore, travel time plays a crucial role in shaping social equity, directly affecting individuals' access to essential services like healthcare (\cite{neutens_accessibility_2015,el-geneidy_cost_2016,chen_understanding_2019}). Unequal travel times may disproportionately impact vulnerable populations who may already face difficulties such as limited mobility and financial constraints, thereby exacerbating systemic inequities. This is particularly noticeable in the case of travel time by public transport, which people of lower income often depend on (\cite{pucher_socioeconomics_2003,jin_examining_2022}).

Travel time has been widely explored in the measurement of accessibility over the years (\cite{osullivan_using_2000,lei_mapping_2010}). In the UK, the Department for Transport (DfT) publishes journey time statistics (\cite{department_for_transport_journey_2019}) which reports travel times of typical journeys from neighbourhood-level small areas known as LSOAs (Lower layer Super Output Areas) to key services in England during the ‘morning peak’ between 7am and 10am, across various travel modes. However, travel time can vary throughout the day for both car and road-based public transport trips due to congestion effects (\cite{weber_bringing_2002,buchel_review_2020}). In the case of public transport, travel times throughout the day are further influenced by factors such as service scheduling (e.g., frequency, interchange) (\cite{ap_sorratini_assessing_2008,buchel_review_2020}), dwell times (e.g., demand-induced boarding and alighting delays) (\cite{lomax_selecting_2003,buchel_review_2020}), etc. The use of a static measurement fails to capture the dynamic nature of travel times for certain geographic areas or types of journeys that do not always occur during morning peak hours. This limitation has become especially evident in the post-COVID era with more diverse journey patterns and more flexible journey times becoming increasingly common (\cite{santana_covid-19_2023}). 

The arrival of General Transit Feed Specification (GTFS) schedule data and various multi-modal routing engines has significantly simplified and accelerated the calculation of origin-destination (OD) based public transport travel times, with various work carried out over the years (\cite{osullivan_using_2000,benenson_measuring_2010,mavoa_gis_2012,dill_predicting_2013,weiss_global_2018}). Despite the clear need for travel time variability to be taken into account when measuring accessibility, most of these works still use a single departure time among origin and destination pairs. Studies have found that travel time variability has a significant impact on accessibility estimates at 50\% on average (\cite{braga_evaluating_2023}). Our work seeks to fill this gap by calculating travel times throughout the day and develop travel time variability metrics to study the geographical inequalities of public transport accessibility.

Previous work has attempted to address this problem. \textcite{lei_mapping_2010} were among the first to recognise the limitation in previous studies which assumes that travel time is the same regardless of time of day. Their subsequent work (\cite{lei_opportunity-based_2012}) highlighted that accessibility by public transport varies throughout the day due to spatial-temporal variations in service levels. They calculated accessibility as the maximum number of opportunities reachable from each census block in Central Los Angeles within pre-defined time buffers (10/20 min). Their analysis compared the results for the morning peak and evening period (after the afternoon peak) and found a significant decrease in accessibility for retail opportunities in the evening. \textcite{fan_impact_2012} calculated job accessibility measures for each hour from 5am to 9pm on weekdays based on travel times for the Minneapolis–Saint Paul Twin Cities Metropolitan Area. These time specific accessibility measures were then aggregated into a single weighted average accessibility measure for the day, which unfortunately diminished the impact of travel time variability on accessibility measurement. To address this, \textcite{owen_modeling_2015} proposed a time-continuous transit accessibility where they take into account trips with varying departure times at every minute between 7am and 9am. They subsequently compared the maximum and average accessibility during this period, and found that average accessibility is almost always lower than the maximum. Furthermore, they calculated various measures of variance for accessibility during the chosen period and found that city centre areas with high frequency public transport services experienced low levels of variability in accessibility levels during the morning peak hours, whereas the suburbs experienced high variability of accessibility. However, their work is still focused on a specific time of day.

What perhaps most closely resembles our proposed research method is the work carried out by \textcite{farber_temporal_2014}. In their study, they calculated travel times from every census block in Cincinnati, Ohio, to the nearest supermarkets at different times of day. Specifically, they were able to calculate travel times for departures at one-minute intervals over a 24-hour period using public transport schedule data. Using these results, they computed several easily interpretable summary statistics, including average and standard deviation, to measure accessibility. Their analysis further explored variations in accessibility levels among different demographic groups, including race, income, and age, highlighting inequalities across these populations.

Our study aims to calculate travel times by bus from every LSOA in England to the nearest healthcare facilities, including hospitals and general practitioners (GPs), on an hourly basis throughout the day. This has been made possible by the Bus Open Data Service (BODS) (\cite{department_for_transport_bus_2020}), which provides comprehensive open-source bus schedule data for England for this study. Doing this will enable us to develop metrics to quantify bus travel time variability to assess the spatial-temporal variability and inequality in accessibility to healthcare from a new dynamic perspective, which has been gaining attention recently (\cite{chen_measuring_2017,chen_understanding_2019,bimpou_dynamic_2020}). Recent years have also witnessed an increased effort to ``level up" the UK, with a drive to reduce the strong geographical inequalities across the country. Transport has always been a strong focus of this initiative as it provides people with access to more opportunities. As such, the government has pledged to improve public transport services nationwide to match the standards of London. Our data-driven approach will provide strong evidence-based insight on precisely where such improvements are most needed. We focus on evaluating bus travel as it is the most prevalent, and often the only mode of public transport in many parts of the country.

We specifically chose healthcare trips because they are more likely to occur at any time of day during working hours, making travel time variability a critical factor in shaping accessibility to these essential services. Accessibility to healthcare facilities has been widely studied at regional (\cite{langford_multi-modal_2016}) and global scales (\cite{weiss_global_2020}) using GIS-based approaches. Studies have shown that public transport plays a significant role in accessibility to healthcare facilities particularly in rural areas in the UK (\cite{haynes_potential_2003}), where areas with better levels of bus services not only experiencing shorter average estimated travel times to the nearest GP, but also having access to more choices of GPs.

By conducting our research, we seek to answer two sets of research questions:

\begin{enumerate}[label=\arabic*.]
    \item How does bus travel time variability vary geographically and how does this contribute to the inequalities of access to healthcare across the country?
    \item What is the relationship between travel time variability and deprivation? Do more deprived areas also experience higher levels of travel time variability?
\end{enumerate}

\section{Method} 

\subsection{Data}
Our analysis uses bus timetable data, road network data, administrative boundary data, healthcare facility location data, and deprivation data.

Our analysis relies on accurate travel times calculations from origins to destinations. To achieve this, we use a routing engine which requires two types of datasets, namely public transport schedule data (in GTFS format) and road network data from OpenStreetMap (OSM). We retrieve bus schedule data from the UK's Bus Open Data Service (BODS) (\cite{department_for_transport_bus_2020}) in GTFS format (General Transit Feed Specification), which contain all information necessary to calculate travel times such as the stops, routes, trips, and schedules of buses. Managed by the UK's Department for Transport (DfT), BODS is the first service of its kind in the world to provide accurate, real-time, and open access public transport data at a large national scale, covering the entirety of England. Since 2020, it has been a legal requirement (via the Bus Services Act 2017) for all bus operators in England to publish their timetables, routes, and other service information (fare, disruption, etc.) to BODS, which ensures the datasets completeness in terms of coverage. All datasets published through BODS are required to meet compliance standards defined by the Department for Transport, ensuring a baseline level of data quality and consistency. Road network data were obtained using Python package \textit{PyDriosm} (\cite{fu_pydriosm_2023}) in PBF (Protocol Buffer Binary) format, and were used to calculate walking time to, from, or between stops when routing with public transport.

We used Lower layer Super Output Areas (LSOAs) as the basic unit of analysis for our study. LSOA is a geographic hierarchy used in the UK Census for reporting small area statistics. In total, there are 33,755 LSOAs in England, with each LSOA typically housing 1000–3000 residents (400–1200 households). Travel times were calculated from each LSOA in England to the nearest hospitals and GPs. We obtained LSOA boundary data alongside other spatial data from the Open Geography portal of the UK's Office for National Statistics (ONS). Hospital and GP location data were obtained from NHS England Digital (\cite{nhs_england_digital_estates_2024,nhs_england_digital_patients_2024}) which are the same sources used by the DfT's official Journey Time Statistics (2019) (\cite{department_for_transport_journey_2019}). We obtained the 2024 version of the two datasets for our study. For hospitals, we applied the same filtering criteria as detailed in the technical report of the DfT Journey Time Statistics (\cite{department_for_transport_journey_2019}) to include only the larger (by floor space) NHS general hospitals in our destination dataset. The resulting dataset contains 222 hospitals and 6308 GPs for year 2024, as opposed to 219 hospitals and 6866 GPs in the 2019 destinations dataset.

For the analyses comparing travel time variability to deprivation, we used data from the English Indices of deprivation 2019 (\cite{ministry_of_housing_communities__local_government_english_2019}). The Index of Multiple Deprivation is a measure if relative levels of deprivation in the UK based on various dimensions of deprivation including income, employment, education, health, crime, housing, and living environment. It is an important tool for understanding and addressing inequalities.

\subsection{Calculating travel time}

We calculate bus travel time between every LSOA and their nearest hospitals and GPs using R\textsuperscript{5}py (v0.1.1), a Python implementation of the R\textsuperscript{5} routing engine providing rapid realistic routing on multimodal transport networks (\cite{fink_r5py_2022}). R\textsuperscript{5} is widely used in both academic and policy contexts. It has been widely validated and adopted in both academic and policy applications for scheduled-based transport accessibility analysis (\cite{verduzco_torres_public_2024,fink_travel_2024,graff_constructing_2024,maciejewska_when_2025}). Compared with some of its competitors such as OpenTripPlanner 1, R\textsuperscript{5} is said to be much faster and less memory intensive (\cite{opentripplanner_2_comparing_2023}). Moreover, R\textsuperscript{5} is highly optimised to capture variations in travel time across time windows and account for uncertainty in waiting times on frequency-based routes, which makes it more applicable for our use case.

\begin{figure}[H]
    \centering
    \includegraphics[scale=0.5]{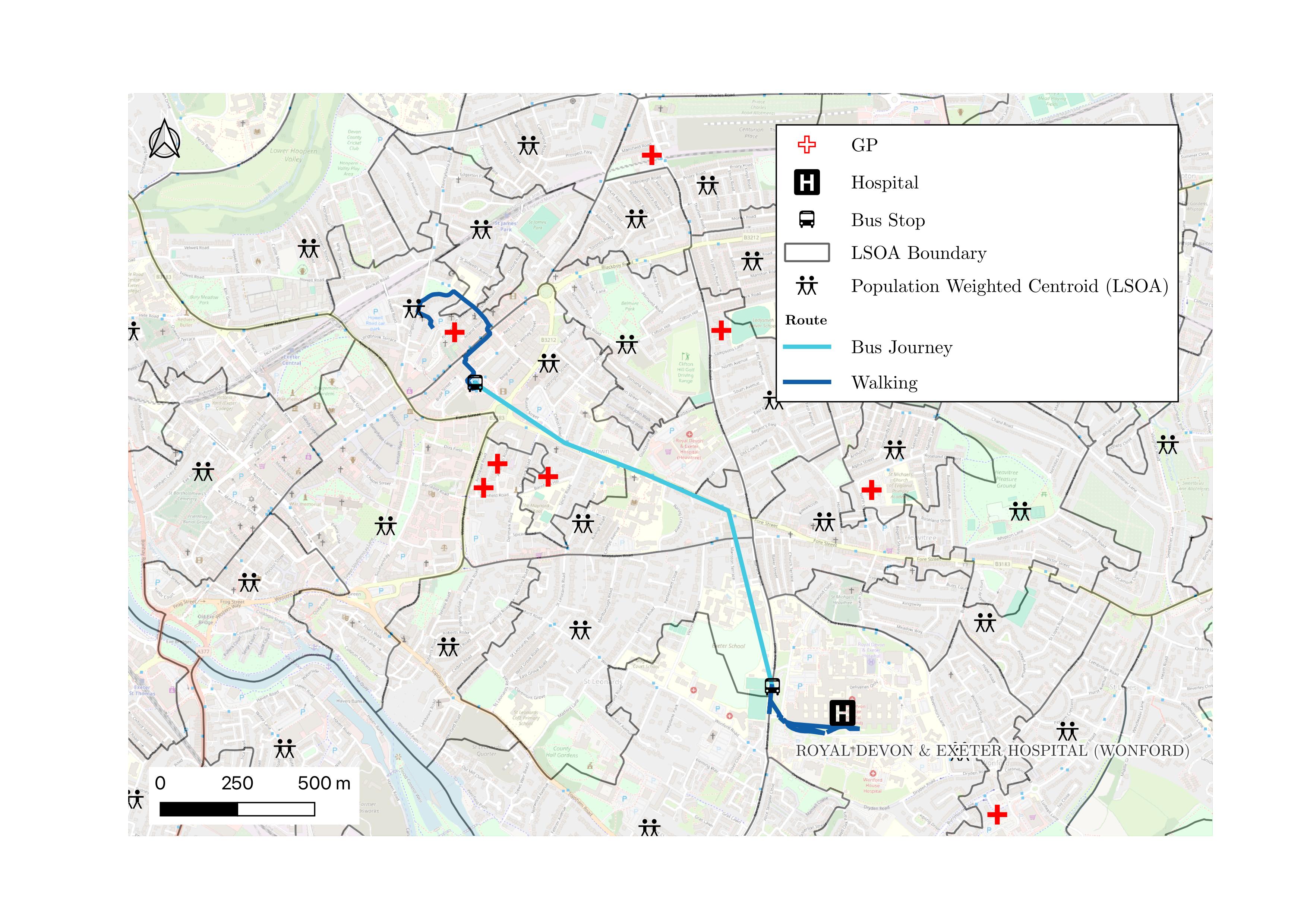}
    \caption{Diagram showing how R\textsuperscript{5} routing engine works. A typical calculated journey begins with a walk from the population weighted centroid of an LSOA (dark blue) to a bus stop, where a bus journey (light blue) starts , and concludes with a walk from the destination bus stop to the final destination.}
    \label{fig:r5}
\end{figure}

We start by building a bus transport network for England using R\textsuperscript{5} which consists of bus routes, schedules, and road networks using the data presented above as input. This network forms an essential part of R\textsuperscript{5}'s \textit{TravelTimeMatrixComputer} with which we can calculate travel times between origin-destination pairs. We initialise the travel time computer using the parameters specified in Table \ref{tab:r5_para}. Note that we use the population weighted centroids of each LSOA as the origin of the trips as it provides a more realistic representation of where people within a geographic area are concentrated, which is crucial for our analysis and for decision-making. It is particularly helpful when analysing rural LSOAs where the geographical centroids might fall in uninhabited locations (e.g., forest, water). Figure \ref{fig:r5} illustrates the routing mechanism of R\textsuperscript{5}. The path from the chosen LSOA to the nearby hospital is colour-coded by segment based on the mode of travel. The journey begins with a walk to the nearby bus stop followed by a bus ride from there, and concludes with a walk to the final destination. A calculated travel time of 10 minutes could  consist of 10 minutes of walking, 10 minutes of riding the bus, or any combination of walking, waiting, and bus riding that totals 10 minutes. In our setting, we do not change the default limits set on the maximum time spent walking (via max\_time\_walking, default 2 hours) nor the number of interchanges allowed (via max\_public\_transport\_rides, default 8) although R\textsuperscript{5} does allows such customisation.

Upon investigation, we further optimised the calculation by aligning trip origins with the nearest bus stops to the centroids. This adjustment emphasises the calculation of travel time specifically on buses, mitigating the variability introduced by differing access times to bus stops within various parts of the LSOA. Furthermore, in many rural areas, bus stops may not be of reasonable walking distance to the majority of the population. Thus, the nearest bus stops provide a more realistic starting point for the majority of trips.

Using the R\textsuperscript{5} travel time computer, we calculate travel times from each LSOA to the nearest hospitals and GPs for departure times scheduled hourly from 09:00 to 17:00 on three selected weekdays in 2024: 30th May, 30th August, and 29th November. These dates were chosen to represent different seasons and to help identify any potential seasonal variations in travel times. We calculate travel times to the five nearest hospitals and GPs, determined by straight-line distance, and use the minimum travel time among them. This approach accounts for scenarios where the closest facilities may not be the quickest to reach by bus. This process results in 9 travel times calculated for each LSOA, one for each hour. We then use these results in our calculation of travel time variability and other metrics.

\begin{table}[H]
    \centering
    \begin{tabular}{|p{4.2cm}|p{8cm}|}
        \hline
        Parameter & Meaning/Value \\ \hline
        transport\_network & Bus transport network for England consisting of bus routes and schedules and road networks \\ \hline
        origins & Population weighted centroid of each LSOA (nearest bus stop) \\ \hline
        destinations & Nearest hospitals and GPs \\ \hline
        departure & Departure time of the trip to be calculated \\ \hline
        departure\_time\_window & 10 min (by default), meaning the engine will compute travel times leaving within 10 minutes after departure \\ \hline
        percentiles & 50 (by default), meaning the final result returned will be the median travel time of all computed trips within the departure time window specified. \\ \hline
        transport\_modes & r5py.TransportMode.TRANSIT, denoting the main mode of transport to use for routing is public transport. \\ \hline
        \raggedright access\_modes \& egress\_modes & r5py.TransportMode.WALK (by default), denoting the mode of transport to and from public transport stops. \\ \hline
        max\_time & 2 hours (by default), denoting the maximum travel time allowed. \\ \hline
        speed\_walking & 3.6 km/h (by default), denoting the walking speed for routing.  \\ \hline
    \end{tabular}
    \caption{Parameters for the R\textsuperscript{5} travel time computer}
    \label{tab:r5_para}
\end{table}

\subsection{Measuring travel time variability and inequality}
We calculate average travel time and travel time variability (TTV) for each LSOA. These measures are then aggregated at a higher administrative level, for which we calculate an inequality measure which quantifies its internal inequality in terms of average travel time and TTV.

We define and calculate the TTV metric as the standard deviation of the travel times across different departure times. We chose standard deviation as it is a well-established statistical measure for dispersion that is both easy to compute and interpret. Despite it lacking the ability of capturing the direction of variability, it is still valuable in a study like ours that examines a broader picture of TTV across a large geographic area. Although standard deviation can be sensitive to outliers, in our case, this works to our advantage. For example, some LSOAs are underserved by buses only at certain times of day, particularly during off-peak hours, leading to significantly longer travel times. Standard deviation effectively amplifies this effect in its value which reflects the effect of service regularity in our TTV metric. Ultimately, we hope the TTV metric we create contributes towards the measurement of overall service dependability from the user's perspective.

Understanding the regional inequalities of travel times and their variability is crucial in our study. In England, Local Transport Authorities (LTA), which are typically embedded within combined authorities (CA) and county councils (CC) or unitary authorities (UA), have oversight of public transport within their regions. They work with bus operators to ensure public transport meets local needs. Therefore, we aggregate the travel time metrics at LAD level and also calculate the inequality metrics (via the Gini coefficient) for all LADs. Originally designed to capture income inequality (\cite{gini_measure_1936}), the Gini coefficient has evolved to measure other forms of inequality. In the context of our study, the Gini coefficient allows us to quantify the internal inequality within LADs regarding average travel times and their variability, providing a complementary perspective to TTV. The Gini coefficient for an LAD is calculated as follows:
\begin{equation}
G = \frac{\sum_{i=1}^{n} (2i - n - 1) \cdot x_i}{n \cdot \sum_{i=1}^{n} x_i}
\end{equation}
where $n$ denotes The total number of LSOAs, $x_i$ denotes The $i$-th value in the sorted array of values ($x_1, x_2, \ldots, x_n$) of average travel times and TTVs, and $\sum_{i=1}^n x_i$ denotes the sum of all values. This formula quantifies the overall dispersion of average travel times and TTV within the LAD. Gini coefficient can range from 0 to 1. A Gini coefficient of 0 indicates perfect equality (all LSOAs in an LAD experience the same average travel times throughout the day and the same TTV). As the Gini coefficient approaches 1, it signals increasing inequality. Some LSOAs within the LAD may have consistently good access to healthcare (low TTV), while others face more variable travel times (higher TTV). In this context, a high Gini coefficient closer to 1 indicates high internal inequalities in healthcare accessibility, which may point to areas where targeted policy intervention is most needed.

\subsection{Statistical analysis}
We apply spatial autocorrelation analysis to try to understand the spatial patterns (e.g., clustering or dispersion) of travel time and its variability. We also perform correlation analysis between the calculated travel time and inequality metrics.

We first perform a global spatial autocorrelation analysis which evaluates the overall spatial pattern across the entire study area (i.e., England). It indicates whether TTV exhibit a clustered, dispersed, or random spatial distribution using \textbf{Moran's I} statistic, which is calculated as follows:
\begin{equation}
I = \frac{n \sum_{i=1}^n \sum_{j=1}^n w_{ij} (x_i - \bar{x})(x_j - \bar{x})}{\sum_{i=1}^n (x_i - \bar{x})^2 \cdot \sum_{i=1}^n \sum_{j=1}^n w_{ij}}
\end{equation}
where $n$ denotes the total number of spatial units (i.e., LSOAs), $x_i$ and $x_j$ denote the values of the variable at locations $i$ and $j$, respectively, $\bar{x}$ denotes the mean, of the variable, $w_{ij}$ denotes the spatial weight between locations $i$ and $j$, typically based on adjacency (e.g., Queen, Rook) or distance (set threshold or k-nearest neighbour). In our analysis, we use k-nearest neighbour ($k = 10$) to construct the weight matrix for all LSOAs. This is because some LSOAs are surrounded by LSOAs which cannot reach healthcare facilities within 2 hours on the bus, making them `islands', which in the case of contingency-based spatial weight, would have no neighbours. We selected $k = 10$ as a balance between ensuring that each LSOA had a sufficient number of neighbours to detect meaningful spatial patterns, while avoiding overly diffuse spatial relationships that could obscure local structure. $k = 10$ provided the best balance between capturing local spatial structure and detecting spatial outliers, particularly high-low (HL) clusters. These outliers are of particular interest for policymakers, as they highlight areas with unusually poor accessibility (in terms of TTV) compared to their neighbours and thus signal potential targets for intervention. In contrast, lower values such as $k = 5$ failed to reveal such patterns reliably. By choosing $k = 10$, the analysis is better able to detect local inequalities in accessibility, like places where TTV are much worse than in surrounding areas. These are exactly the kinds of inequalities that matter for planning transport and healthcare services fairly.

We subsequently carried out local spatial autocorrelation analysis which focuses on specific locations, identifying areas where clustering or dispersion occurs using local indicators of spatial association (LISA). We calculated \textbf{Local Moran’s I} for each LSOA as follows:
\begin{equation}
I_i = \frac{(x_i - \bar{x})}{S^2} \sum_{j=1}^n w_{ij} (x_j - \bar{x})
\end{equation}
where $I_i$ denotes the Local Moran’s I statistic for LSOA $i$ and $S^2$ denotes the variance of $x$. Significant values of Local Moran’s I reveal spatial clusters of high-highs (high values surrounded by high values) or low-lows (low values surrounded by low values) or spatial outliers of high-low (high values surrounded by low values) or low-high (low values surrounded by high values) relationships.

For both global Moran’s I and local indicators of spatial association (LISA), we used 999 random permutations to assess statistical significance. A standard significance threshold of $p<0.05$ was applied to identify areas with statistically significant spatial autocorrelation. For LISA, only clusters meeting this threshold were interpreted as meaningful spatial patterns. This approach ensures that the spatial clustering results are robust and not due to random chance.

In addition to spatial autocorrelation analysis, we performed a series of correlation analysis using the Pearson correlation coefficient to examine the linear relationships between pairs of average travel time, travel time variability, and the inequality metrics of both (via the Gini coefficient). To account for the potential inflation of false positives due to multiple testing, we applied the False Discovery Rate (FDR) correction using the Benjamini-Hochberg procedure to adjust all resulting $p-values$.

\section{Results}
In this section, we report the results of our analysis of travel times and their variability.

\subsection{Travel time variability at a national scale}
We begin by presenting visualisations of the calculated travel time variability (TTV) metric at a national scale. Before doing so, we first compare the TTV results across the three selected dates (30th May, 30th August, and 29th November). We calculate pairwise Pearson correlation coefficient between the distribution of TTVs of the three dates. High correlation in TTV distributions across the three dates is observed (Table \ref{tab:seasonal_corr}), suggesting that the travel time variability patterns are consistent over time. Therefore, we focus on reporting the results for 30th May, though visualisations for the other two dates are provided in the appendix for readers' reference.

\begin{table}[h!]
    \centering
    \begin{tabular}{|l|lll|lll|}
\hline
\multirow{2}{*}{} & \multicolumn{3}{l|}{Hospital}                                            & \multicolumn{3}{l|}{GP}                                                  \\ \cline{2-7} 
                  & \multicolumn{1}{l|}{30th May} & \multicolumn{1}{l|}{30th Aug} & 29th Nov & \multicolumn{1}{l|}{30th May} & \multicolumn{1}{l|}{30th Aug} & 29th Nov \\ \hline
30th May          & \multicolumn{1}{l|}{1.00}     & \multicolumn{1}{l|}{0.94}     & 0.87     & \multicolumn{1}{l|}{1.00}     & \multicolumn{1}{l|}{0.95}     & 0.90     \\ \hline
30th Aug          & \multicolumn{1}{l|}{0.94}     & \multicolumn{1}{l|}{1.00}     & 0.89     & \multicolumn{1}{l|}{0.95}     & \multicolumn{1}{l|}{1.00}     & 0.93     \\ \hline
29th Aug          & \multicolumn{1}{l|}{0.87}     & \multicolumn{1}{l|}{0.89}     & 1.00     & \multicolumn{1}{l|}{0.90}     & \multicolumn{1}{l|}{0.93}     & 1.00     \\ \hline
    \end{tabular}
    \caption{\textbf{Correlation matrix} showing pairwise correlations between TTVs of the three selected dates in May (30th), August (30th), and November (29th). The matrix shows overwhelmingly high correlations between TTV distributions throughout the year, indicating that seasonal variations of theoretical TTV is low.}
    \label{tab:seasonal_corr}
\end{table}

\begin{figure}[H]
    \centering
    \begin{subfigure}[b]{\textwidth}
        \centering
        \includegraphics[width=0.8\linewidth]{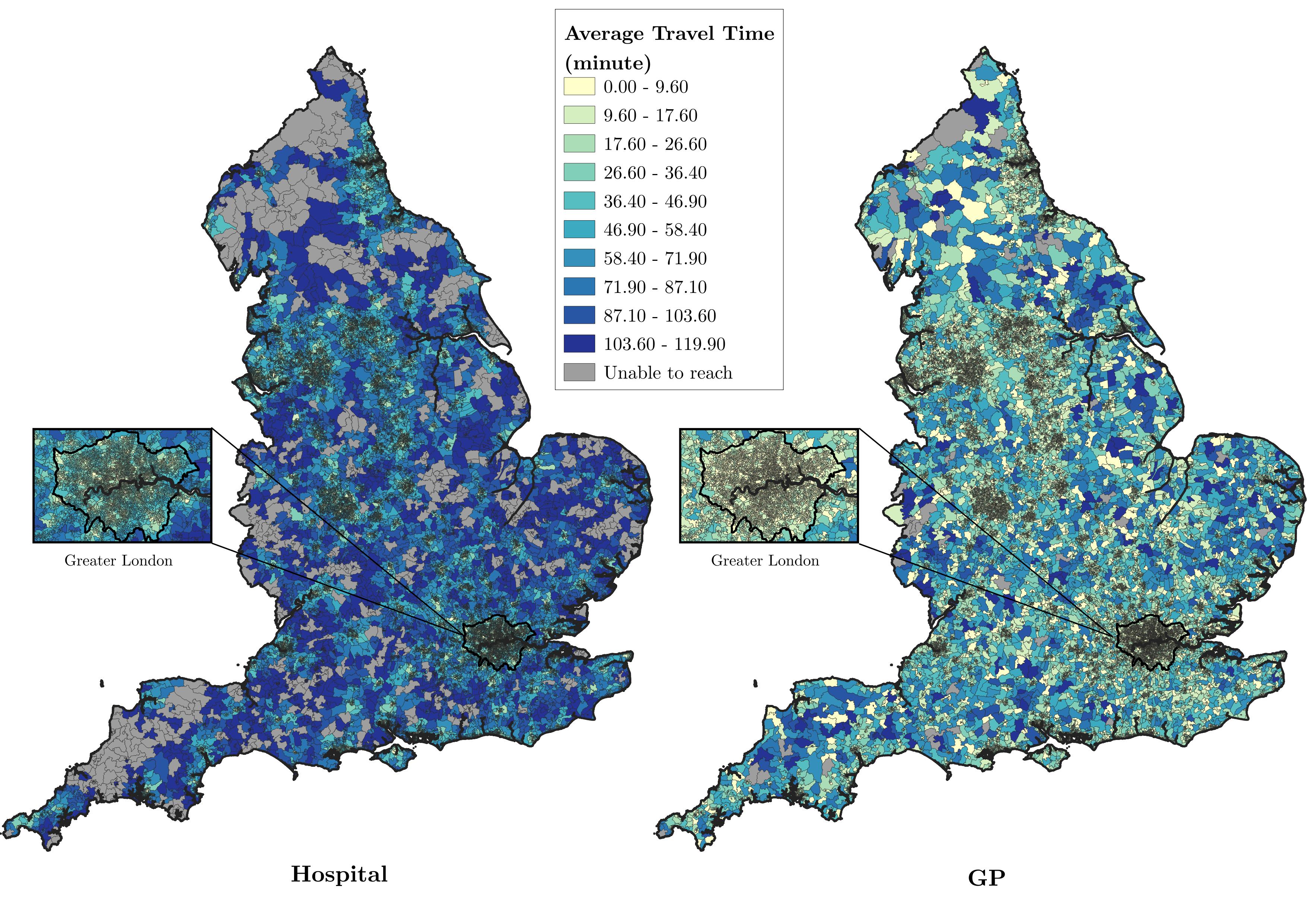}
        \caption{Average travel time from LSOAs in England to hospitals and GPs}
        \label{fig:hosp_gp_mean}
    \end{subfigure}
    \hfill 
    \begin{subfigure}[b]{\textwidth}
        \centering
        \includegraphics[width=0.8\linewidth]{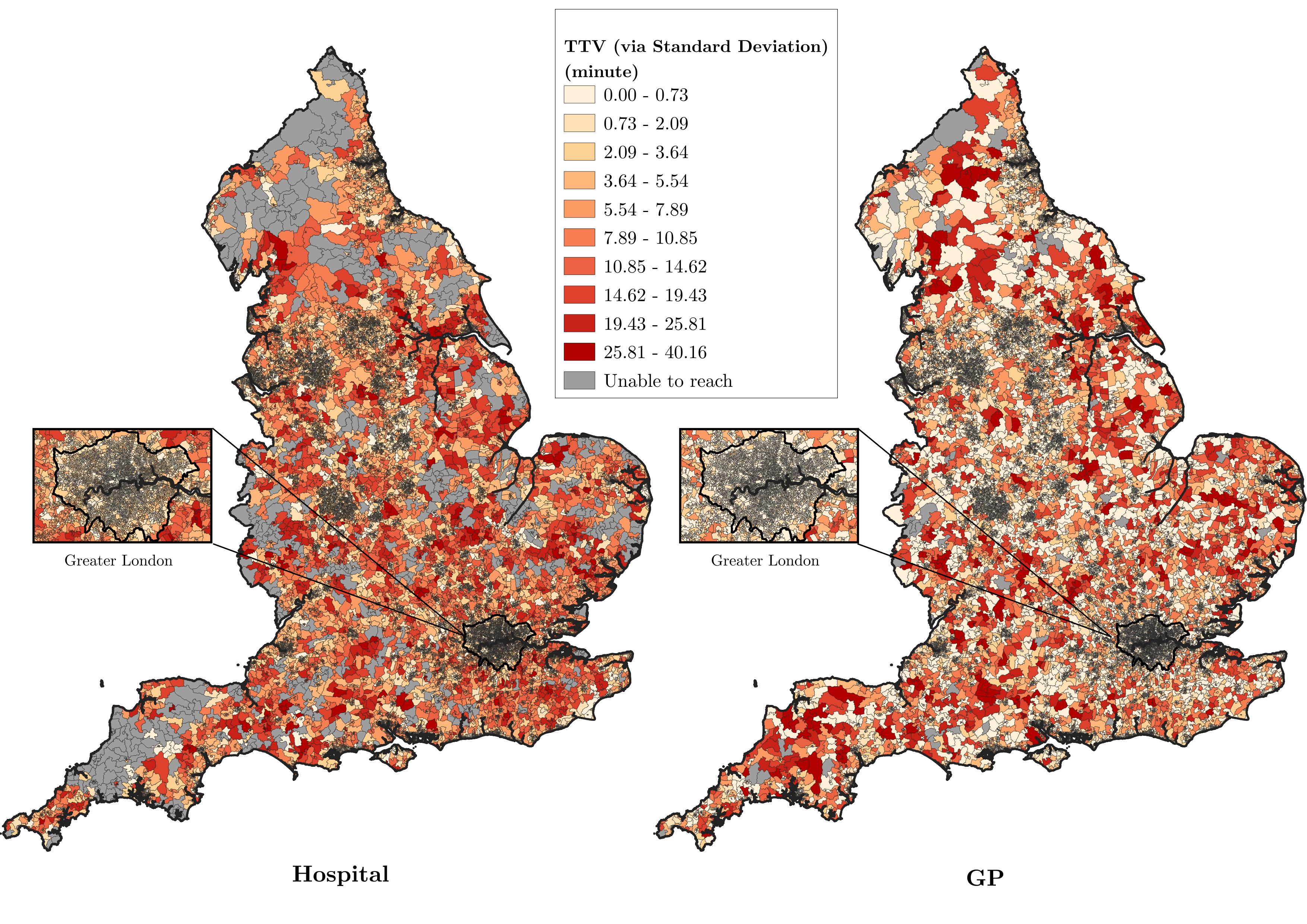}
        \caption{Travel time variability from LSOAs in England to hospitals and GPs}
        \label{fig:hosp_gp_sd}
    \end{subfigure}
    \caption{Maps of LSOAs in England showing (a) average travel time and (b) travel time variability (TTV) to hospitals and GPs on 30th May.}
    \label{fig:ttv}
\end{figure}

Figure \ref{fig:ttv} depicts the average travel times (\ref{fig:hosp_gp_mean}) and travel time variability (\ref{fig:hosp_gp_sd}) to hospitals and GPs across England on 30th May. Although these maps do not capture all regional nuances or explain causal mechanisms, they provide an important overarching perspective on healthcare accessibility across the country. Overall, average travel times to hospitals are longer than that to GPs. The spatial distribution of average travel times and TTV to GPs and hospitals are a reflection of the distribution of the healthcare facilities themselves. GPs are far greater in quantity than hospitals and more widely distributed within local communities to provide convenient primary care access. They are strategically located to serve smaller catchment areas, ensuring accessibility within a short distance for most people. In terms of TTV, Figure \ref{fig:hosp_gp_sd} visualises TTV using standard deviation of travel time across the day. Overall, a greater number of LSOAs exhibit smaller TTV to GPs compared to hospitals, reflecting once again the higher quantity and greater density of GP practices. The maps also show clusters of LSOAs in the South West and the North of England that cannot reach the nearest hospital by bus within 2 hours while no clusters of comparable size were observed for GPs. Across the country, there are far more LSOAs that cannot reach their nearest hospitals (708 out of 33755) than those that cannot reach the nearest GPs (52 out of 33755).

While Figure \ref{fig:ttv} provides a descriptive overview to contextualise the spatial patterns of accessibility and its variability, we can gain further insights into the spatial patterns of TTV across England by conducting a spatial autocorrelation analysis. We begin with a global spatial autocorrelation analysis on TTV for LSOAs in England. The results, as shown in Table \ref{tab:global_sa} measured using Moran's I statistics, reveal a relatively strong and statistically significant positive spatial autocorrelation for TTV for access to hospitals, while a weaker yet still evident positive spatial autocorrelation is observed for access to GPs. This indicates that areas with similar levels of TTV tend to cluster geographically rather than being randomly distributed. High TTV values are more likely to be surrounded by other high values, and low TTV values tend to cluster with other low values. This suggests the presence of spatially structured inequalities in travel time variability, reflecting regional disparities in healthcare accessibility.

\begin{table}[h!]
    \centering
    \begin{tabular}{|c|c|c|}
        \hline
         & Hospital & GP \\ \hline
        Moran's I & 0.57 & 0.25 \\ \hline
        p-value & 0.001 & 0.001 \\ \hline
        z-score & 243.72 & 106.82 \\ \hline
    \end{tabular}
    \caption{Global Moran's I statistics for travel time variability (TTV) from LSOAs in England to hospitals and GPs}
    \label{tab:global_sa}
\end{table}

\begin{figure}[htbp]
    \centering
    \includegraphics[scale=0.5]{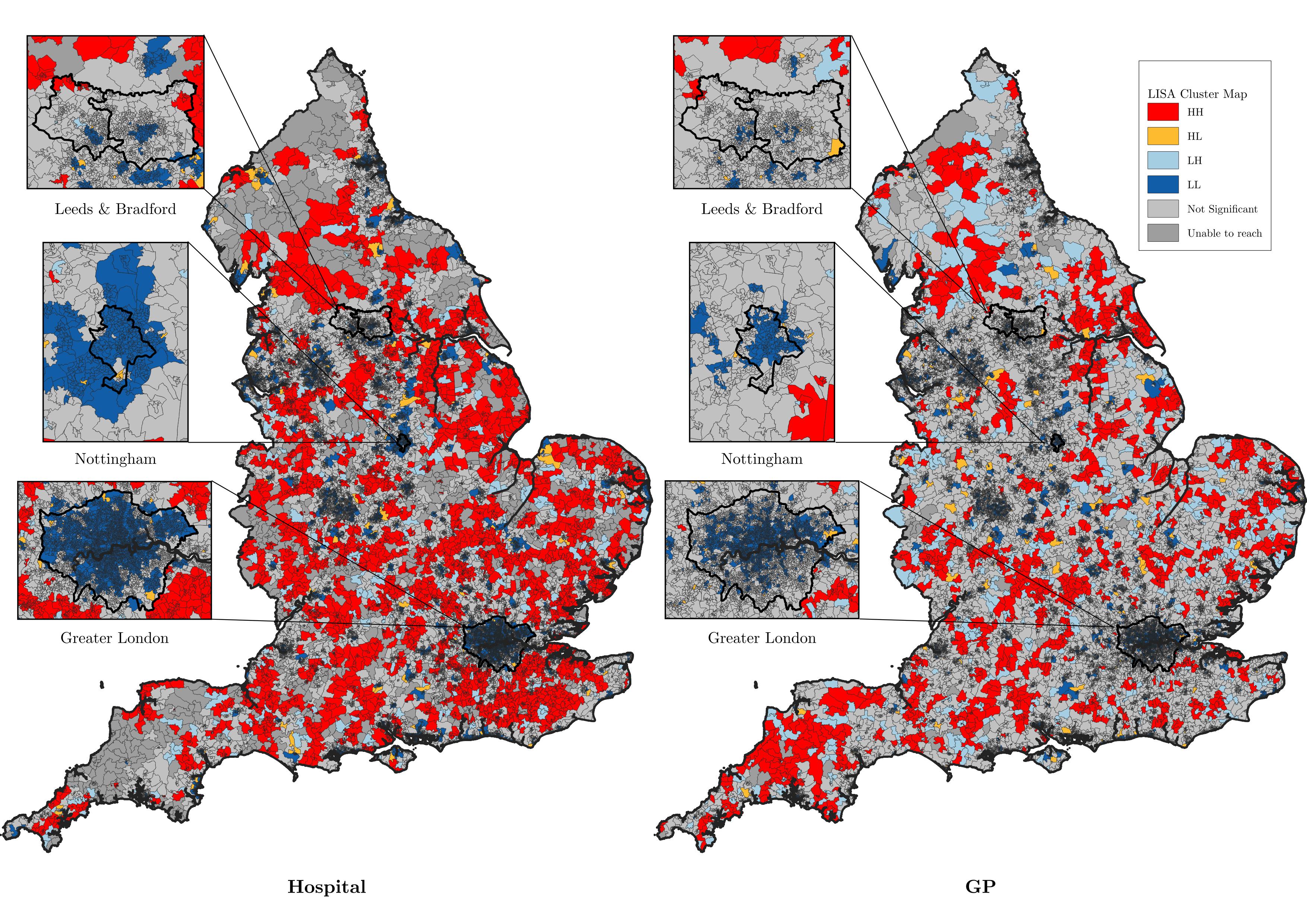}
    \caption{\textbf{Local Indicators of Spatial Association (LISA)} clustering map for travel time variability (TTV) to hospitals and GPs from LSOAs in England. \textbf{High-high clusters} (red) are mainly observed in rural areas across the country, while \textbf{low-low clusters} (dark blue) are concentrated in major urban centres (see zoomed in maps of selected metropolitan areas). Additionally, \textbf{high-low outliers} (yellow) and \textbf{low-high outliers} (light blue) are present, though they are less visually prominent on a map of this scale. These spatial patterns highlight the geographical inequalities of TTV, with urban areas generally exhibiting more consistent travel times and rural areas experiencing greater variability.}
    \label{fig:lisa}
\end{figure}

While the global spatial autocorrelation results provide an overall measure of clustering, local spatial autocorrelation analysis using LISA (Local Indicators of Spatial Association) allows us to identify specific areas where significant spatial patterns occur. The LISA plots in Figure \ref{fig:lisa} reveal distinct spatial clusters and outliers of TTV across England. High-high clusters, which are areas where both an LSOA and its neighbouring ones exhibit high TTVs, are observed throughout the country, typically in rural areas. In contrast, low-low clusters, where both an LSOA and its neighbours have low TTV values, are primarily observed in major urban areas, reflecting the consistent and regular level of bus service provision in these regions. The City of Nottingham exhibits an extensive low-low clusters of TTV for hospital access covering almost the entire city boundary and some surrounding areas. This coincides with the fact that Nottingham has arguably the best bus service outside London in England (\cite{nottingham_city_transport_nottinghams_2024}). Meanwhile, cities with comparable or larger size like Leeds and Bradford exhibits much smaller low-low clusters covering only their core city centre areas. It is worth mentioning that Leeds is the largest city in England without a metro system and therefore bus transport is the absolute backbone of its entire public transport system. This heavy dependence on a single mode of public transport system may exacerbate the poor access to healthcare compared to cities like Nottingham, where alternative means of transport (e.g., a modern tram system) could help mitigate such issues. Greater London exhibits the largest cluster of LSOAs with low TTV (low-low clusters). Interestingly, high-high clusters are observed in areas immediately outside the Greater London boundary, particularly for access to hospitals (as shown in the zoomed in map in Figure \ref{fig:lisa}). These areas typically consists of commuter towns with a significant proportion of their population working in London. These areas may have good public transport links with Central London, but lack good local public transport to key amenities, which results in the observed pattern.

\begin{figure}[htbp]
    \centering
    \includegraphics[scale=0.45]{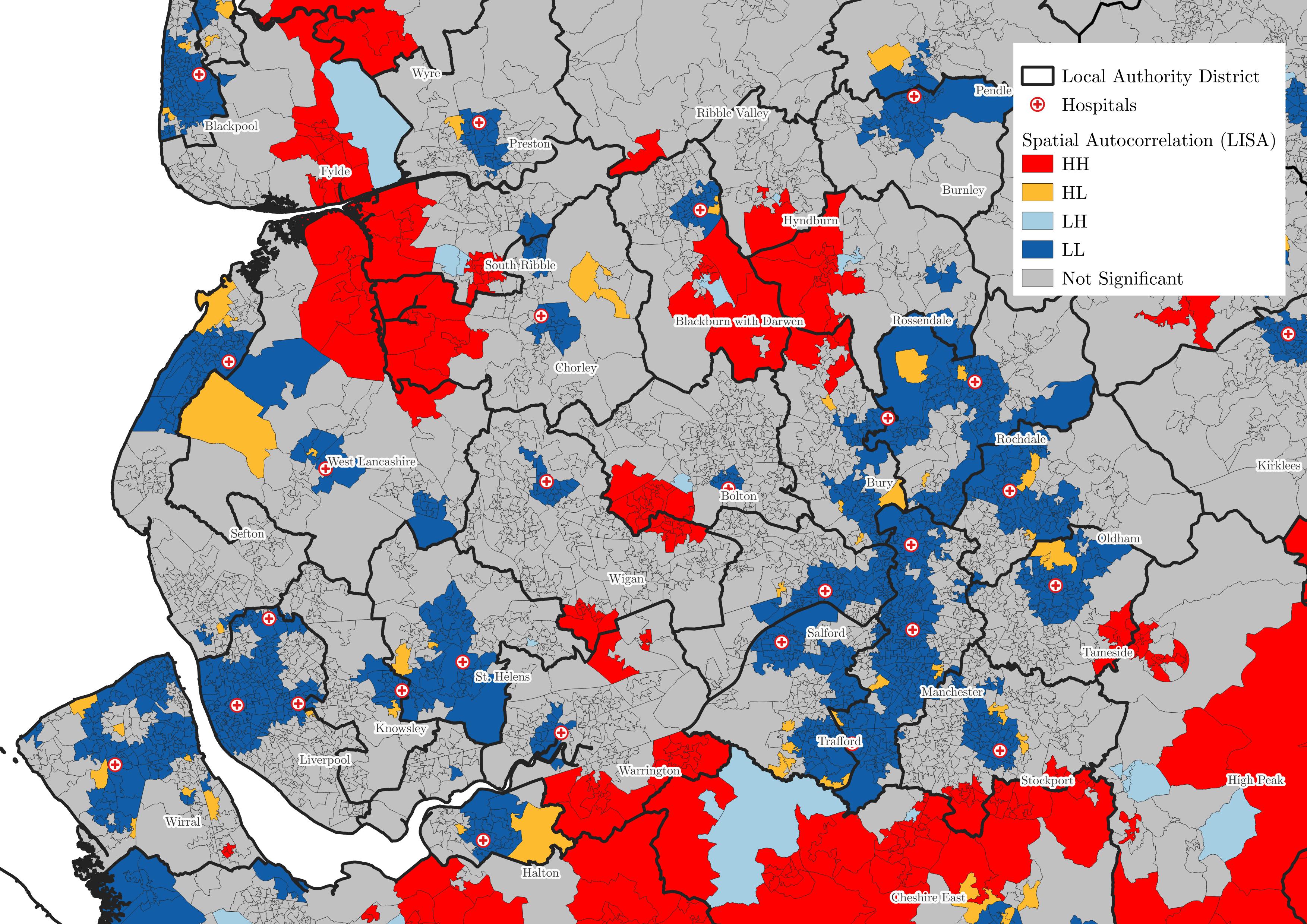}
    \caption{\textbf{Local Indicators of Spatial Association (LISA)} clustering map for travel time variability (TTV) to hospitals from LSOAs in a large metropolitan area in North West England consisiting of Liverpool and Manchester. \textbf{High-low outliers} (yellow) can be clearly observed, suggesting these particular LSOAs experience significantly higher TTV than the surrounding LSOAs.}
    \label{fig:lm_lisa}
\end{figure}

Spatial outliers, such as high-low patterns, are also of great interest not only from an analytical standpoint but also for stakeholders such as local policymakers and public transport operators. Figure \ref{fig:lm_lisa} shows the LISA map for a large combined metropolitan area consisting of parts of Liverpool City Region and Greater Manchester, home to nearly 5 million residents. High-low outliers (coloured in yellow) can be observed in LSOAs in the region, indicating these particular LSOAs experience significantly higher TTVs than their surrounding LSOAs. This pattern may be due to localised factors such as gaps in service frequency and built environment factors and almost always has to be examined on a case by case basis. While our results help identify these priority regions that may benefit from targeted policy interventions to reduce travel time variability, further analysis is needed to uncover the root causes of these patterns and to design effective, targeted interventions. This would require additional data such as local built environment characteristics (e.g., street connectivity, infrastructure, and land use, etc.). Such insights would enable local transport authorities to develop more tailored strategies to reduce TTV and improve equitable healthcare access.

\begin{figure}[htbp]
    \centering
    \includegraphics[width=\linewidth]{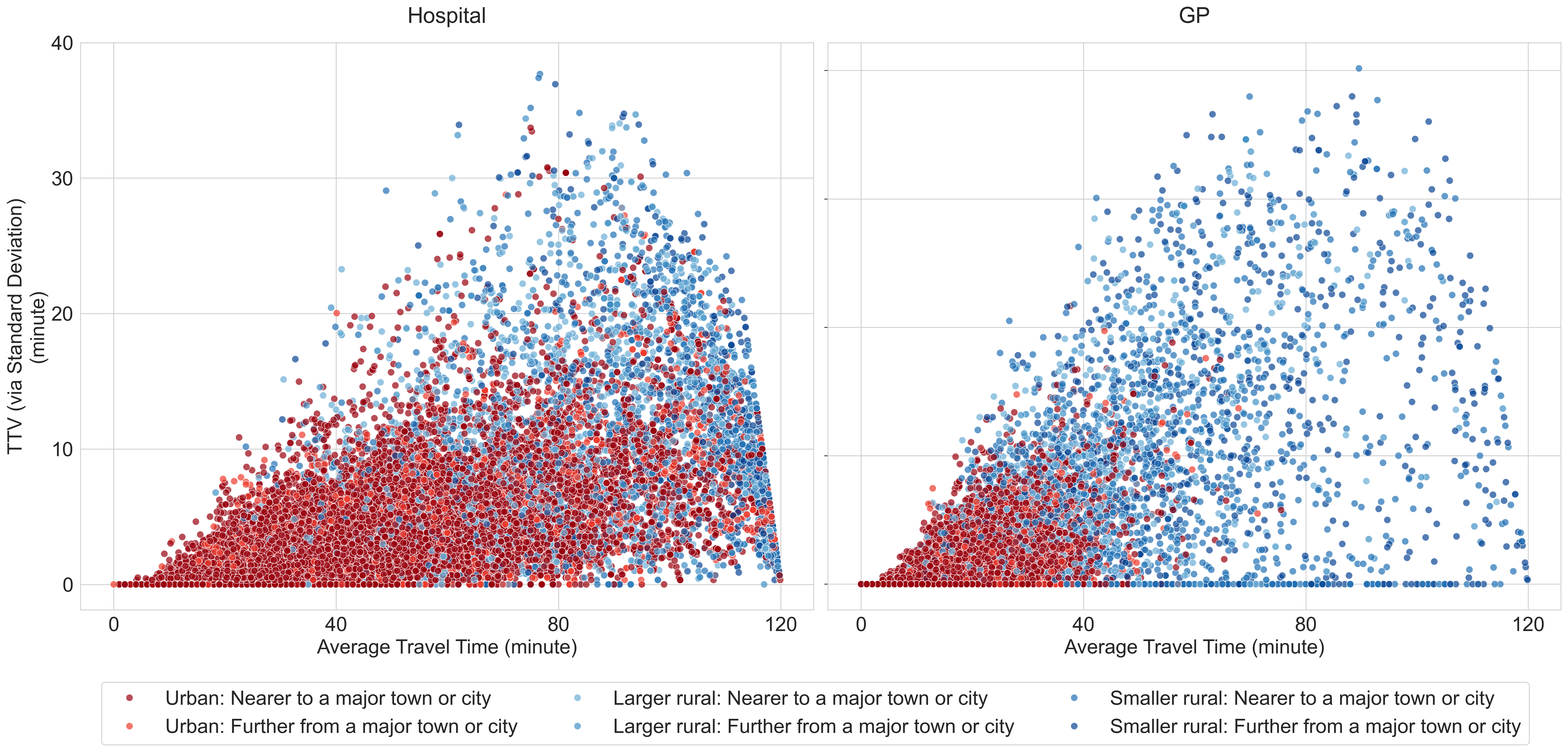}
    \caption{\textbf{Scatterplot} of travel time variability (TTV) versus average travel time for trips to hospitals and GPs from each LSOA in England, with points color-coded by settlement type (urban/rural). The plot reveals a general positive correlation between TTV and average travel time, with urban areas exhibiting lower values for both metrics, while rural areas tend to show higher variability and longer travel times. This pattern is less distinct for hospital access with some urban areas exhibiting high average travel times while certain rural areas displaying low TTV, suggesting access to hospitals is influenced by additional factors beyond the rural-urban classification, including the limited number of hospitals and their uneven geographical distribution.Overall, the distribution highlights the urban-rural divide in accessibility from both dimensions.}
    \label{fig:rural_urban}
\end{figure}

Building on the observed urban-rural gap in healthcare access revealed by the spatial autocorrelation analysis, we now explicitly examine the differences in TTV between rural and urban areas across the country. This allows us to assess how TTV varies across settlement type and whether we can identify any potential systemic disparities. Figure \ref{fig:rural_urban} shows that, for access to GPs, the majority of the urban LSOAs are scattered around the bottom left corner of the plot, indicating low average travel time and low TTV. On the other hand, rural LSOAs tend to have larger average travel time and TTV. For hospitals, while the same clear divide does not entirely hold, a similar trend can still be observed. 
Many urban LSOAs are concentrated toward the lower left corner of the plot, indicating relatively low average travel times and TTV. However, the distribution is more spread out compared to that of access to GPs, with some urban LSOAs exhibiting higher average travel times and TTV. Similarly, rural LSOAs tend to have larger average travel times and TTV, but the pattern is less distinct, with lots of overlap between rural and urban areas. This suggests that access to hospitals is influenced by additional factors beyond the rural-urban classification, including the significantly fewer number of hospitals and the geographical distribution of them. It is also worth noting that there is a considerable number of rural LSOAs with very high average travel time whilst having low TTV for access to hospitals. This could be considered the least desirable scenario, since it indicates that people living in these LSOAs experience consistently long travel times; on the other hand, having high average travel time and high TTV indicates that, at certain times of the day, travel times are shorter so people can try to plan their journeys accordingly.

\begin{figure}[H]
    \centering

    \begin{subfigure}[b]{\textwidth}
        \centering
        \includegraphics[width=0.99\linewidth]{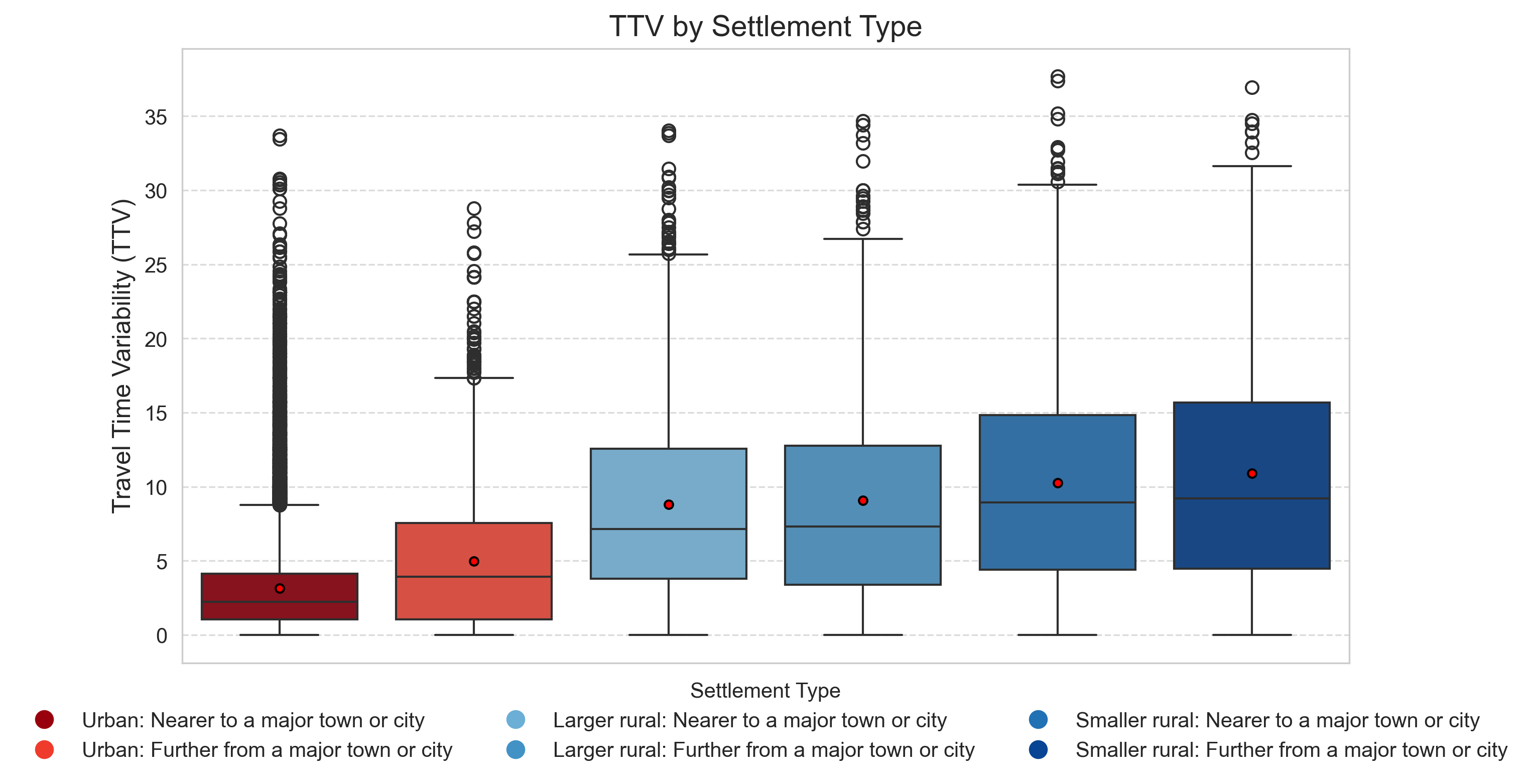}
        \caption{TTV for hospital access}
        \label{fig:ttv_by_settlement_hosp}
    \end{subfigure}
    \hfill 
    \begin{subfigure}[b]{\textwidth}
        \centering
        \includegraphics[width=0.99\linewidth]{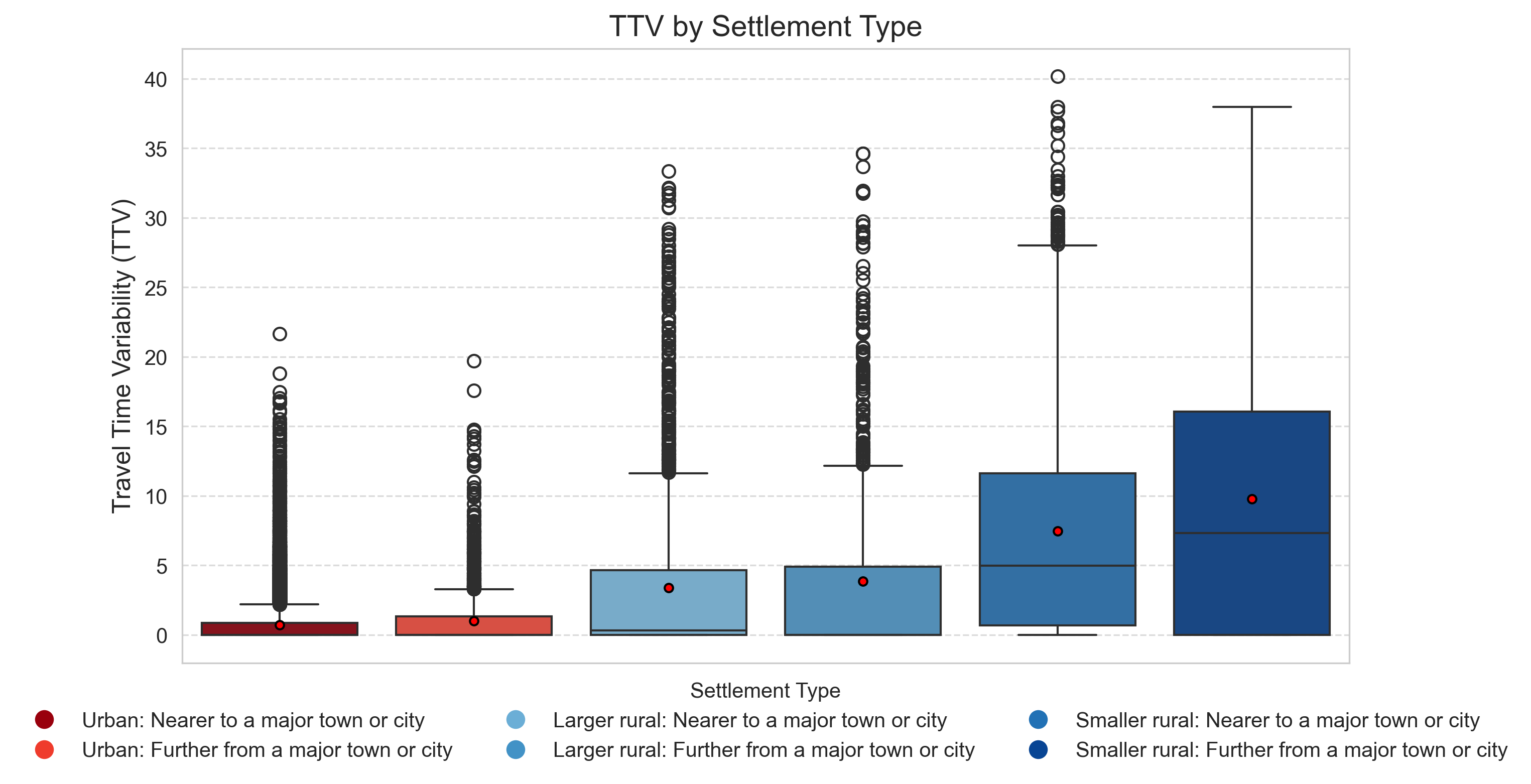}
        \caption{TTV for GP access}
        \label{fig:ttv_by_settlement_gp}
    \end{subfigure}
    
    \caption{\textbf{Boxplots} of the distribution of TTV by different settlement and destination types. The plot reveals that as areas become more rural, the median and mean TTV increases. In particular, predominantly rural areas show not only higher medians and means but also greater interquartile ranges.}
    \label{fig:ttv_by_settlement}
\end{figure}

Table \ref{tab:summary_stats} shows summary statistics of TTV for hospital and GP access, divided by binary settlement types. To complement the visual trends shown in Figure \ref{fig:rural_urban} and address the urban-rural divide more precisely, we include boxplots displaying the distribution of TTV by different settlement and destination types (Figure \ref{fig:ttv_by_settlement}). The settlement types are ordered from most urban (nearer to a major town or city) to small rural (further from a major town or city), using the official The 2021 Rural Urban Classification (\cite{office_for_national_statistics_2021_2025}). The plot reveals a consistent pattern: as areas become more rural, the median and mean TTV increases. In particular, predominantly rural areas show not only higher medians and means but also greater interquartile ranges, indicating both higher and more variable TTV. This provides evidence  that rural communities face greater unpredictability in public transport access to healthcare, reinforcing concerns about service availability and regularity in less densely populated areas.

\begin{table}[H]
    \centering
    \begin{tabular}{|l|ll|ll|}
\hline
\multirow{2}{*}{}  & \multicolumn{2}{l|}{Hospital}      & \multicolumn{2}{l|}{GP}            \\ \cline{2-5} 
                   & \multicolumn{1}{l|}{Urban} & Rural & \multicolumn{1}{l|}{Urban} & Rural \\ \hline
Count (number of LSOAs)              & \multicolumn{1}{l|}{28134} & 4913  & \multicolumn{1}{l|}{28199} & 5504  \\ \hline
Mean (minute)               & \multicolumn{1}{l|}{3.29}  & 9.57  & \multicolumn{1}{l|}{0.76}  & 5.80  \\ \hline
Median (minute)             & \multicolumn{1}{l|}{2.30}  & 8.09  & \multicolumn{1}{l|}{0.00}  & 2.07  \\ \hline
Standard Deviation (minute) & \multicolumn{1}{l|}{3.41}  & 7.06  & \multicolumn{1}{l|}{1.53}  & 7.87  \\ \hline
IQR (minute)                & \multicolumn{1}{l|}{3.29}  & 9.72  & \multicolumn{1}{l|}{0.93}  & 9.21  \\ \hline
Min (minute)                & \multicolumn{1}{l|}{0.00}  & 0.00  & \multicolumn{1}{l|}{0.00}  & 0.00  \\ \hline
Max (minute)                & \multicolumn{1}{l|}{33.71} & 37.68 & \multicolumn{1}{l|}{21.66} & 40.16 \\ \hline
    \end{tabular}
    \caption{Summary statistics of TTV for hospital and GP access, divided by settlement type (binary).}
    \label{tab:summary_stats}
\end{table}

\subsection{Travel time variability at a regional scale}
So far, our analysis uses LSOAs as unit of analysis, which is a very fine local scale. In England, Local Transport Authorities (LTA), which are typically embedded within combined authorities (CA) and county councils (CC) or unitary authorities (UA), have oversight of public transport within their regions. They work with bus operators to ensure public transport meets local needs. To better understand the geographic inequalities in TTV at a broader regional level, we examine TTV aggregated at a higher administrative level, specifically Local Authority Districts (LAD). Although the mapping between LADs and LTAs does not perfectly align (e.g., several LADs jointly form a CC or CA that overseas public transport for a larger area), we believe this aggregation is appropriate for examining TTV at a suitable regional level.

\begin{figure}[H]
    \centering
    \includegraphics[width=\linewidth]{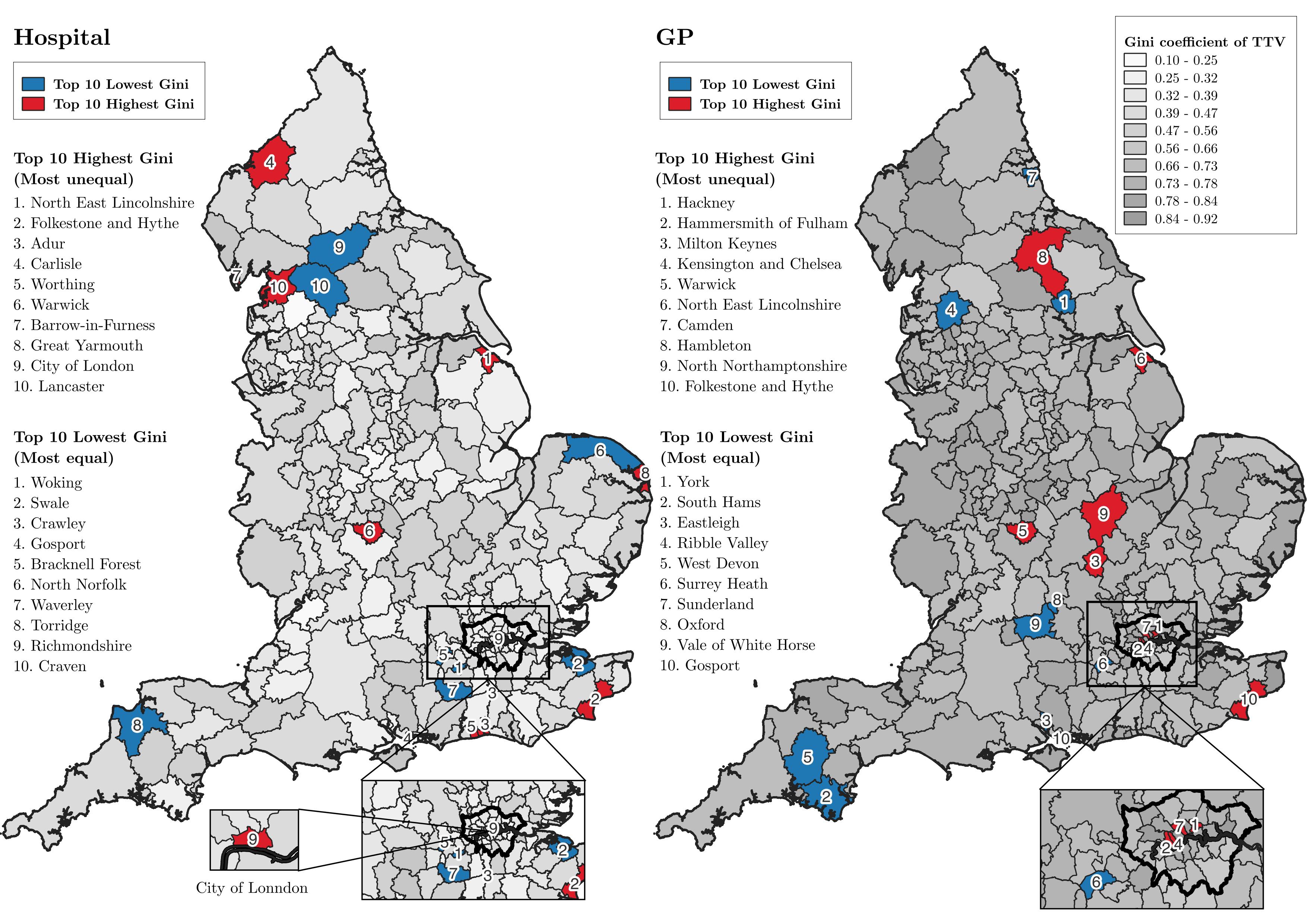}
    \caption{Maps showing the Gini coefficient of TTV for each LAD, with the ten most unequal LADs highlighted in red and the ten most equal in blue.}
    \label{fig:gini_ttv_lad}
\end{figure}

We aggregate travel time metrics at the Local Authority District (LAD) level by computing the mean travel time and TTV across all LSOAs within each LAD. Additionally, we calculate inequality measures, specifically the Gini coefficient, for both travel time and TTV in each LAD. Figure \ref{fig:gini_ttv_lad} shows the Gini coefficient of TTV for each LAD, with the ten most unequal LADs in TTV highlighted in red and the ten most equal in blue. Several observations can be drawn from the map. Firstly, Gini coefficients for GP access are generally higher than those for hospitals, indicating that LADs are more unequal in terms of TTV to GPs than to hospitals within themselves. However, lower inequality for TTV to hospitals is not necessarily positive, as it may reflect uniformly high TTVs across an LAD, particularly in largely rural areas, rather than equitable and quick access. For hospital access, six of the ten LADs with the lowest Gini coefficients are located in one of the `home counties' of Surrey, Berkshire, Hampshire, and Kent, suggesting better equity in accessibility to hospitals across towns within London’s commuter belt. Interestingly, no London boroughs appear among the ten LADs with lowest Gini. In fact, much to our surprise, the City of London ranks among the top ten highest Gini (most unequal). This suggests that while London may have overall good public transport with shorter travel times, accessibility levels may not be equal in all areas. This effect is even more pronounced in GP access, where four Inner London boroughs situated just outside Central London, namely Hackney, Hammersmith and Fulham, Kensington and Chelsea, and Camden, are among those with the highest TTV inequality, suggesting unequal access to GP services within these areas in terms of TTV.

\begin{figure}[H]
    \centering
    \includegraphics[width=\linewidth]{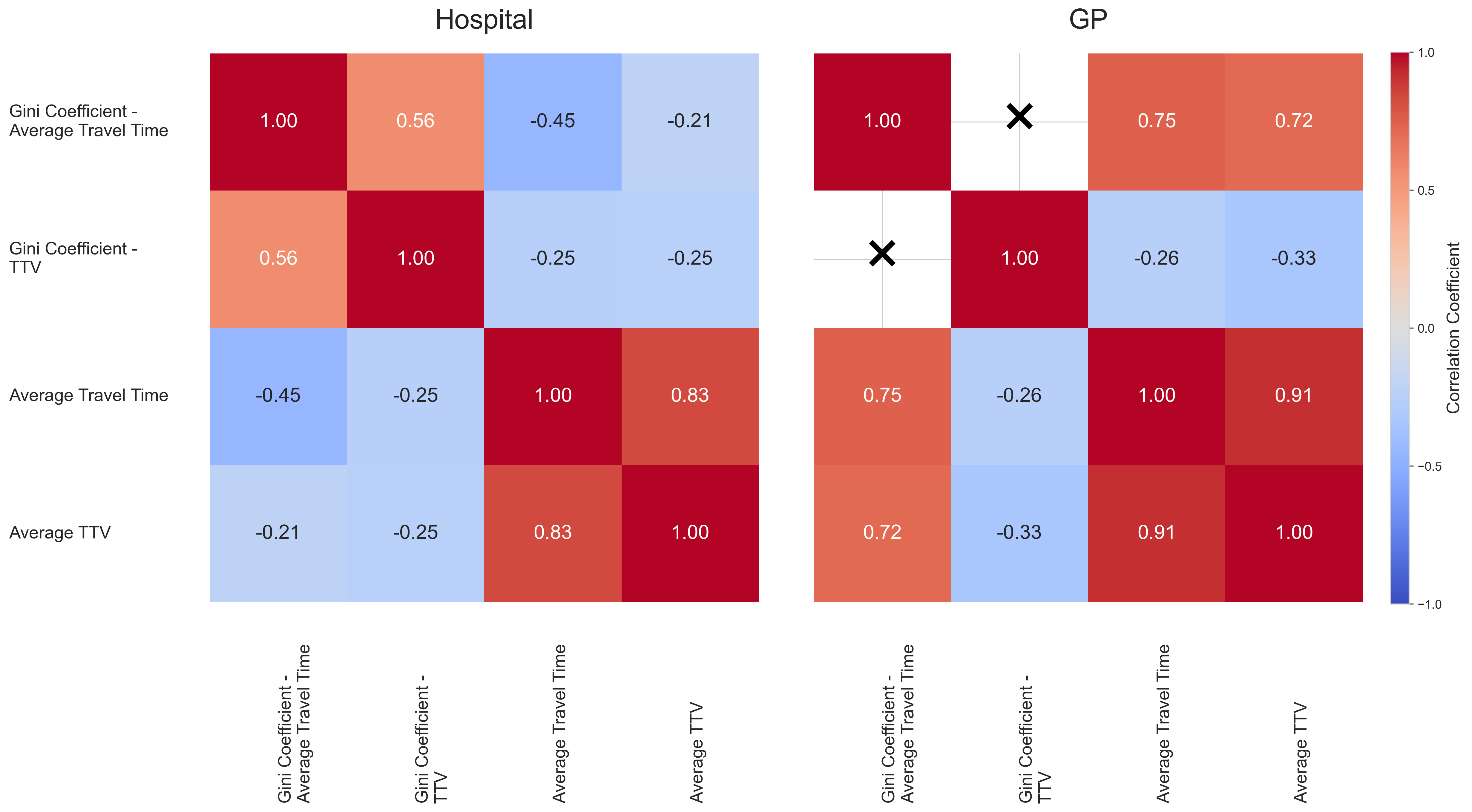}
    \caption{\textbf{Correlation matrix} showing relationships between aggregated average travel time, travel time variability (TTV), and inequality measures (via the Gini coefficient) of them at the Local Authority District (LAD) level. The matrix highlights a much stronger positive correlation between average travel time and TTV at LAD level compared to that at LSOA level. Additionally, some correlations between the Gini coefficients of average travel time and TTV and the aggregated average travel time and TTV were also observed. Interestingly, no significant correlation was observed between the gini coefficient of average travel time and the gini coefficient of TTV for GP access, which has been marked on the matrix as `X'.}
    \label{fig:hosp_gp_corr}
\end{figure}

\begin{figure}[htbp]
    \centering
    \includegraphics[width=\linewidth]{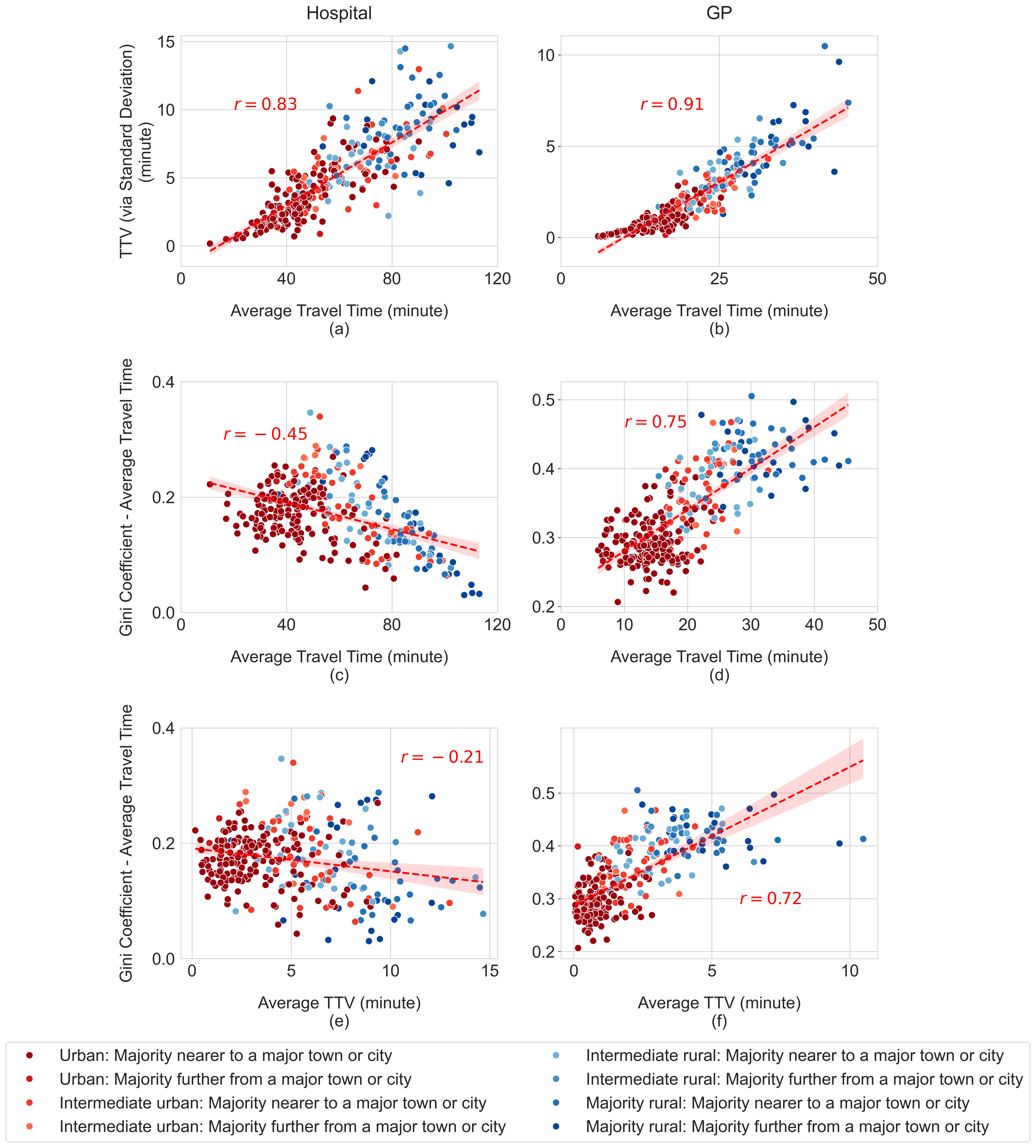}
    \caption{\textbf{Scatterplot} of (a, b) travel time variability (TTV) versus average travel time, both aggregated at Local Authority District (LAD) level; (c, d) Gini coefficient of average travel time at LAD level versus aggregated average travel time at LAD level; (e, f) Gini coefficient of average travel time at LAD level versus aggregated average TTV at LAD level. Points are color-coded by settlement type (urban/rural).}
    \label{fig:lad_corr}
\end{figure}

Figure \ref{fig:hosp_gp_corr} depicts the results of the correlation analysis between the average travel time and average TTV of each LAD, as well as the calculated Gini coefficients as inequality metrics. We observe strong positive correlations between aggregated average travel time and TTV at the LAD level for both access to hospitals (\( r = 0.76 \)) and GPs (\( r = 0.9 \)), unlike the results for individual LSOAs (as shown in Figure \ref{fig:rural_urban}) which do not exhibit a strong linear relationship.  These results suggest a much clearer linear relationship when examining travel time and TTV at the aggregated regional level. Figure \ref{fig:lad_corr}a and  \ref{fig:lad_corr}b depicts the linear relationships where the points are colour coded according to the type of LAD (rural/urban). A similar pattern to that at LSOA levels can be observed where urban LADs tend to have lower average travel time and average TTV while rural LADs have higher. It is also worth noting that urban LADs with significant rural areas (coloured in light red) sit perfectly in the middle ground, exhibiting higher average travel times and average TTVs than most other LADs that are more urban, but lower than those of LADs that are largely or mainly rural.

The correlation analysis results as shown in Figure \ref{fig:hosp_gp_corr} also reveals correlation between the inequality metrics and average travel time and average TTV for LADs. For access to GPs, strong correlations were observed between the inequality of average travel times and the average travel times and average TTV. Figure \ref{fig:lad_corr}d and \ref{fig:lad_corr}f shows scatter plots where such correlation can be observed. The points representing each individual LAD are again colour coded to indicate whether it is urban or rural. LADs that are more urban tend to also be less unequal in average travel times among LSOAs in them. These positive correlations indicate that LADs that have higher average travel time and higher TTV to GPs also tend to be more unequal in average travel times in themselves. This implies that LADs with longer average travel times tend to exhibit greater inequality in accessibility, meaning that some areas within these LADs experience much longer travel times than others. Meanwhile, LADs with greater inequality in travel times also tend to have higher TTV which leads to less predictable travel times, potentially exacerbating access challenges for certain populations.

We also examined the same relationship between the inequality of average travel times and the average travel times and average TTV for access to hospitals. Unexpectedly, we found a moderate negative correlation between the Gini coefficient of average travel times and the average travel time (\( r = -0.45 \)) and a very weak negative correlation between the Gini coefficient of average travel times and the average TTV (\( r = -0.21 \)) as shown in Figure \ref{fig:lad_corr}c and \ref{fig:lad_corr}e. This indicates that, as the average travel time increases, the inequality in travel times tends to decrease; which in turn suggests that, in LADs with longer average travel times, the distribution of travel times is more equal, meaning that residents in these areas experience relatively similar travel times regardless of their location (either uniformly high or uniformly low). This is likely due to the spatial distribution of hospitals. In rural LADs with a small number of hospitals, only a few LSOAs may have relatively short travel times to these facilities, while the majority of the population experiences uniformly long travel times. This uniformity reduces the inequality measure (as most travel times are similarly high), even though the average travel time is also high.

\subsection{Travel time variability and deprivation}

Our analysis also examined the relationship between travel time variability and deprivation. TTV can disproportionately affect people living in deprived areas where people are less likely to own cars and access to more reliable public transport options is often limited. Specifically, we wanted to examine whether deprived areas are also more likely to experience higher TTV. It is worth noting that the latest deprivation statistics for England at LSOA level, know as the English indices of deprivation (2019) (\cite{ministry_of_housing_communities__local_government_english_2019}) used LSOAs defined by the 2011 UK Census. However, our TTV calculation is based on LSOA boundaries defined in the latest 2021 UK Census. Between the two censuses, most LSOA boundaries remained the same, while some were split and some were merged. We therefore use the official ONS LSOA (2011) to LSOA (2021) Exact Fit Lookup (\cite{office_for_national_statistics_lsoa_2024}) to match the two iterations of LSOAs, to enable us to examine the correlation between TTV and IMD.

\begin{figure}[H]
    \centering
    \includegraphics[width=\linewidth]{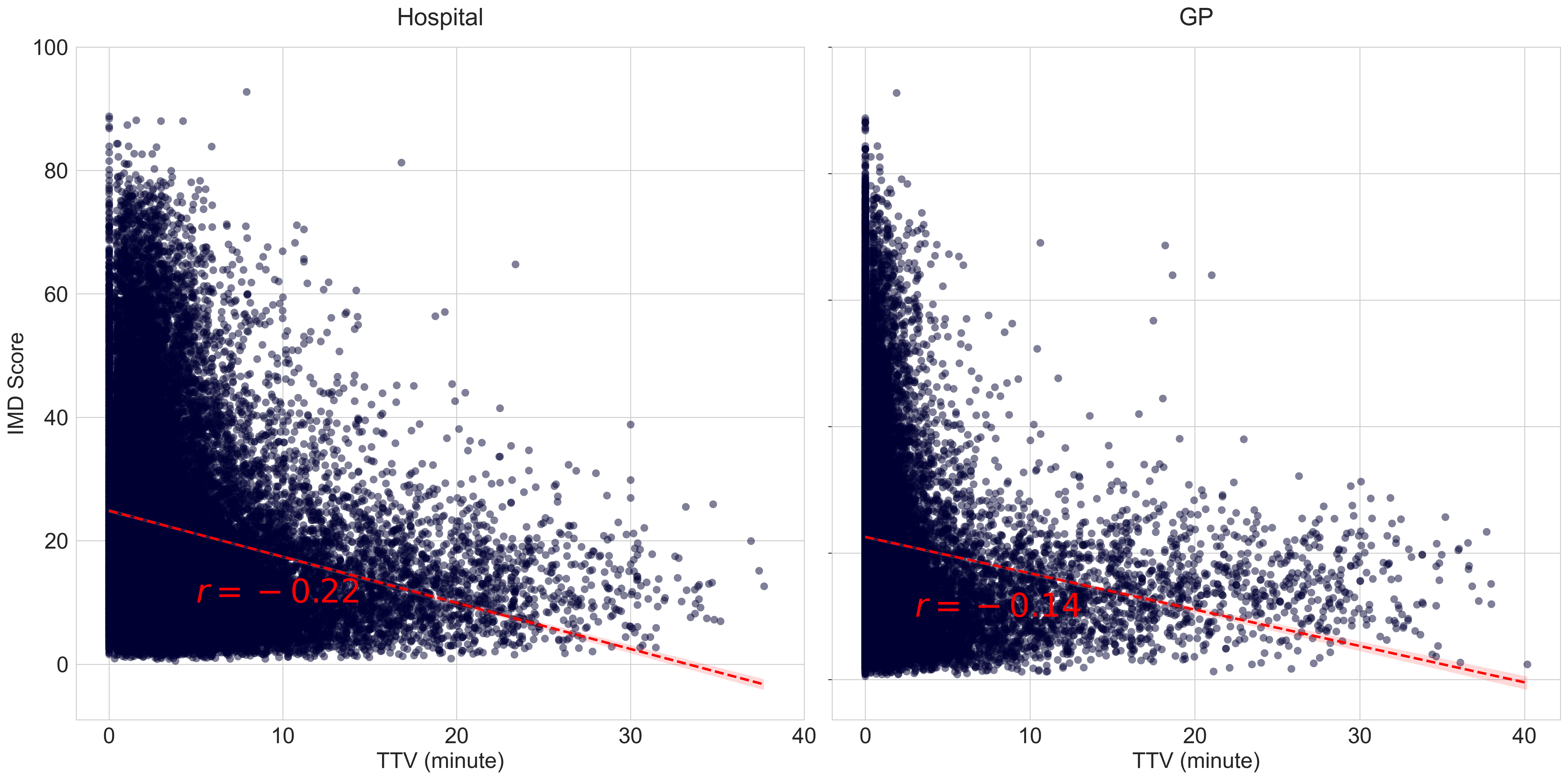}
    \caption{\textbf{Scatterplot} of the IMD Score (0$\sim$100) versus travel time variability (TTV, in minute). A weak negative correlaton is observed, suggesting that more deprived areas (i.e. higher IMD Score) tend to have lower TTV. This suggests a complex relationship between TTV and deprivation which may partly be attributed to the UK’s unique urban geography. For example, many areas in city centres are often very deprived and simultaneously benefit from well-developed public transport networks leading to lower TTV, and affluent rural areas may be very car dependent with little public transport provision.}
    \label{fig:ttv_dep_corr}
\end{figure}

Using the Pearson correlation coefficient to examine the relationship between TTV and the IMD, we found no explicit correlation between the two variables. Interestingly, we observed an overall downward trend (as shown in Figure \ref{fig:ttv_dep_corr}), suggesting that more deprived areas (i.e. higher IMD Score) tend to have lower TTV. While this finding may initially seem counter-intuitive, it is also unsurprising in the context of the UK’s urban geography. Many city centre areas in the UK often exhibit higher levels of deprivation yet are also well served by public transport, resulting in lower TTVs. This finding provides a degree of reassurance, indicating that deprivation does not always coincide with higher TTV. It also led to our further investigation to identify areas that exhibit both high TTV and high levels of deprivation. Understanding where these overlaps occur is crucial for identifying regions where transport-related inequalities may be exacerbating existing socioeconomic challenges.

\begin{figure}[H]
    \centering
    \includegraphics[scale=0.5]{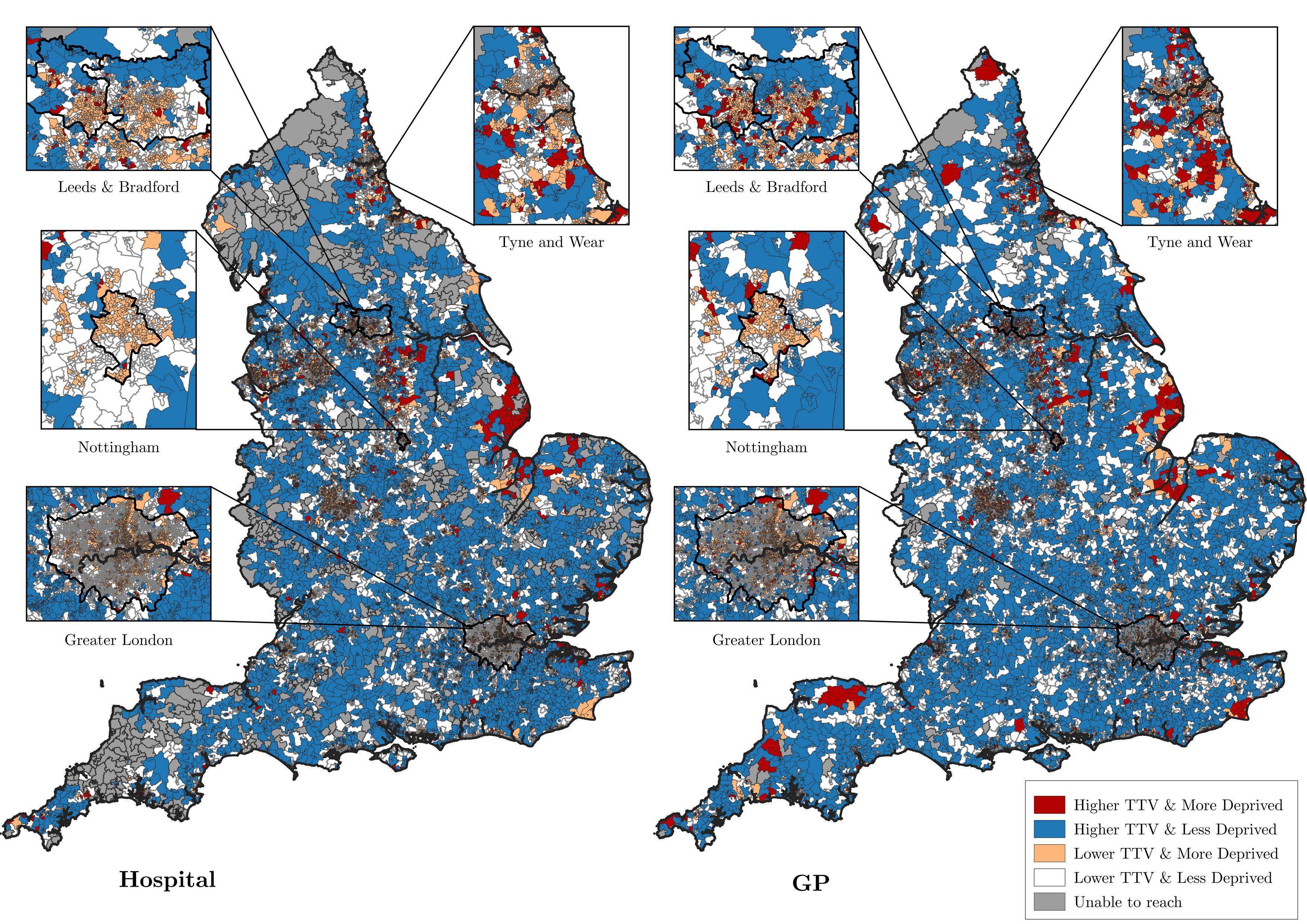}
    \caption{\textbf{Correlation and convergence between TTV and IMD.} LSOAs are colour coded to represent the category they belong to. \textbf{Higher TTV and More Deprived} (red) are the LSOAs ranking in the top 30\% for both TTV and IMD. \textbf{Higher TTV and Less Deprived} (blue) are the LSOAs ranking in the top 30\% for TTV but not for IMD. \textbf{Lower TTV and More Deprived} (peach) are the LSOAs ranking in the top 30\% for IMD but not for TTV. \textbf{Lower TTV and Less Deprived} (white) are the LSOAs that do not fall within the top 30\% for either TTV or IMD.}
    \label{fig:ttv_dep_hosp_gp}
\end{figure}

We therefore categorise LSOAs into four groups based on their distinct characteristics of TTV and deprivation. This classification allows us to better understand the interplay between TTV and deprivation, highlighting areas that may require targeted policy interventions or further investigation. Figure \ref{fig:ttv_dep_hosp_gp} presents a visualisation of the categorisation of the correlation between TTV and the IMD.
\begin{itemize}
    \item \textbf{Higher TTV and More Deprived}: LSOAs in the top 30\% for both TTV and IMD, representing the neighbourhoods with the most pressing concern. Geographically, the map reveals a clear north-south divide for this type of LSOA, with a higher concentration of LSOAs meeting this criteria in the north of England compared to the south. These are the areas that need the most improvement to public transport, as high TTV is likely to exacerbate the existing challenges faced by their residents due to their already disadvantaged socioeconomic status. While these areas are clear priorities for improving public transport access, further analysis is needed to understand the specific local factors contributing to high TTV, such as low demand and incentive to use public transport, worsened by unreliable or irregular services. Addressing these challenges will require more detailed data, including a closer examination of service-level bus timetable data, community health needs assessments, and local mobility patterns gathered by surveys, etc.
    \item \textbf{Higher TTV and Less Deprived}: LSOAs in the top 30\% for TTV, but not for IMD. These areas exhibit relatively high TTV but lower levels of deprivation, suggesting that the high TTV might result from low demand for public transport. These areas (coloured in blue) cover the largest portion of the maps and are typically located in the rural parts of the country. Although these areas are not immediate candidates for equity-driven interventions, they require further investigation especially as transport policy shifts toward net zero goals. Understanding local travel behaviour, vehicle ownership, and travel demand will be essential for designing solutions that improve public transport access.
    \item \textbf{Lower TTV and More Deprived}: LSOAs in the top 30\% for IMD, but not for TTV. These areas are among the most deprived in the country. While they currently exhibit lower levels of TTV, this favourable accessibility may be fragile. Continuous monitoring and further support are needed to maintain this level of service and to meet the demands from future growth and development.
    \item \textbf{Lower TTV and Less Deprived}: LSOAs that are not in the top 30\% for either TTV or IMD. This is an ideal scenario with relatively equitable healthcare access and lower socioeconomic deprivation. Focus should be on maintaining the good levels of public transport provision. However, similar to the above scenario, strategic planning is still necessary to anticipate future shifts in demand, demographics, or service provision. Additional data such as population projections, future transport plans, or planned changes to healthcare provision, could inform proactive strategies to maintain their accessibility advantage.
\end{itemize}

\section{Conclusion}
In our study, we investigated public transport (bus)-based accessibility to healthcare facilities at a small area level across England, focusing on the often-overlooked perspective of travel time variability (TTV). Specifically, we developed a TTV metric for each LSOA and analysed its geographical inequalities across various spatial scales. Additionally, we examined the relationship between TTV and deprivation, identifying distinct scenarios where the interaction of these two metrics could be used to inform targeted policy interventions.
\\

Several observations and conclusions can be drawn from our analysis. Firstly, we attempted to paint a more complete picture of accessibility to healthcare facilities by looking at average daily travel time and TTV together. At the LSOA level, we observed an overall positive correlation between TTV and average travel time for both access to hospitals and GPs. This correlation became much stronger when the metrics were aggregated at the Local Authority District (LAD) level. Our analysis also uncovered a distinct urban-rural divide, with urban areas generally exhibiting shorter average travel times and lower TTV, while rural areas faced longer travel times and higher TTV. However, some distinct pattern were also observed, particularly for hospital access. Notably, a considerable number of urban and rural LSOAs exhibited high average travel times paired with low TTV, indicating consistently high travel times throughout the day. Secondly, the distribution of travel time variability (TTV) at the LSOA level exhibits a clear tendency towards spatial clustering, with areas of high (or low) TTV values often surrounded by similar values. Outliers were also identified where areas differ significantly from their neighbours in TTV, suggesting that those areas may be of particular interest to policy makers for targeted interventions. Thirdly, the examination of the inequalities of travel times and TTV via the calculated Gini coefficient at the LAD level also revealed the distinct characteristics of internal inequalities across different settlement and destination types. Lastly, our study conducted a preliminary analysis of the relationship between TTV and deprivation. Our results indicated no strong correlation between the two, suggesting a complex relationship which may partly be attributed to the UK’s unique urban geography. For example, many areas in city centres are often very deprived and simultaneously benefit from well-developed public transport networks leading to lower TTV, and affluent rural areas may be very car dependent with little public transport provision. We therefore categorised LSOAs into four groups based on their distinct characteristics of TTV and deprivation. This classification provides a foundation for future research aimed at developing targeted interventions tailored to these groups.

It is also important to acknowledge the limitations of our analysis.
First, our analysis only uses bus data, excluding other modes of public transport. While this does not provide a comprehensive view of public transport accessibility, it accurately reflects the current reality for most of the country, particularly in rural areas, where buses are often the only available public transport option. Buses are also often the more economical public transport option and are more frequently chosen for short-distance travel, making them particularly relevant for trips to healthcare facilities.
Second, by using the shortest travel time to healthcare facilities, we make the assumption that individuals will default to using one of the closest options available to them. However, research has shown that this is not always the case. For instance, a study in Eastern England (\cite{haynes_potential_2003}) found that only 56\% of the population were registered with the GP closest to their home. This proportion is lower in urban areas with more provider options and higher in rural areas where choices are limited or accessibility to a broader range of providers is constrained. Individuals’ choices of healthcare facilities are influenced by factors beyond proximity, including cultural norms, social networks, and personal preferences, etc. (\cite{maciejewska_when_2025}). These non-spatial factors often play a more significant role in decision-making than distance and travel time alone. However, such preferences are inherently difficult to quantify and incorporate into accessibility models. 
Future research could address this limitations by adopting the NHS England Digital dataset on GP registrations (\cite{nhs_england_digital_patients_2024}), which includes comprehensive mappings between populations and their registered GPs, enabling more accurate calculations of healthcare accessibility.
Third, although our analysis uses population-weighted centroids for LSOAs, it is important to acknowledge that the impact of high TTV in sparsely populated areas, or areas with low public transport usage, will be smaller. Nevertheless, it is important that even people living sparsely populated areas can benefit from good levels of public transport. Incorporating travel demand and usage in future work could help us fully understand the impact of TTV.
Fourth, our reliance on timetable data means that the TTV metric primarily reflects variability caused by bus service scheduling, which we refer to as theoretical TTV. We acknowledge the significant gap between this theoretical measure and empirical TTV, which could be derived from bus operational data and would account for additional variability caused by operational delays and other real-world disruptions. Nonetheless, our ability to uncover meaningful patterns using timetable data paves way for future research utilising real-time location data. Such studies could provide deeper insights into TTV, potentially uncovering more impactful findings.
Lastly, despite our effort to translate insights gain from the analysis into evidence-based policy recommendations or operational strategies, more granular data would be required to uncover the root causes of the observed patterns and spatial outliers and to design effective, targeted interventions. Access to such data would enable local transport authorities and healthcare planners to develop strategies that are more precisely tailored to local conditions, ultimately helping to reduce TTV and improve equitable access to healthcare.

Overall, our analysis provided a more comprehensive view of accessibility, offering a more robust measure that enhances and complements existing travel time-based accessibility metrics. Our findings laid significant ground for future research on TTV-based accessibility, particularly through the use of real-time data, which could offer a more accurate representation of realistic and realisable accessibility as opposed to the theoretical models based on timetables.

\section{Data Availability}
The bus schedule data used in this study is publicly available through the Bus Open Data Service (BODS) for the most up-to-date bus timetables. However, historical data is not currently publicly available. Data used for this research was downloaded on the dates chosen (30th May 2024, 30th August 2024, and 29th November 2024) and is available via \url{https://doi.org/10.5281/zenodo.15459184}. All other data, including census and the IMD, are publicly available.

\section{Acknowledgement}
This work is funded by the UKRI-EPSRC Centre for Doctoral Training in Environmental Intelligence (ref: EP/S022074/1) through a PhD studentship (grant number: 2717702).

For the purpose of open access, the author has applied a Creative Commons Attribution (CC BY) licence to any Author Accepted Manuscript version arising from this submission.

The authors thank Prof Stewart Barr (University of Exeter) for their valuable input and insights, which improved this work.

\newpage
\printbibliography

\clearpage
\pagestyle{plain}
\appendix
\section{Maps for additional selected dates}
\begin{figure}[H]
    \centering
    \begin{subfigure}[b]{\textwidth}
        \centering
        \includegraphics[width=0.8\linewidth]{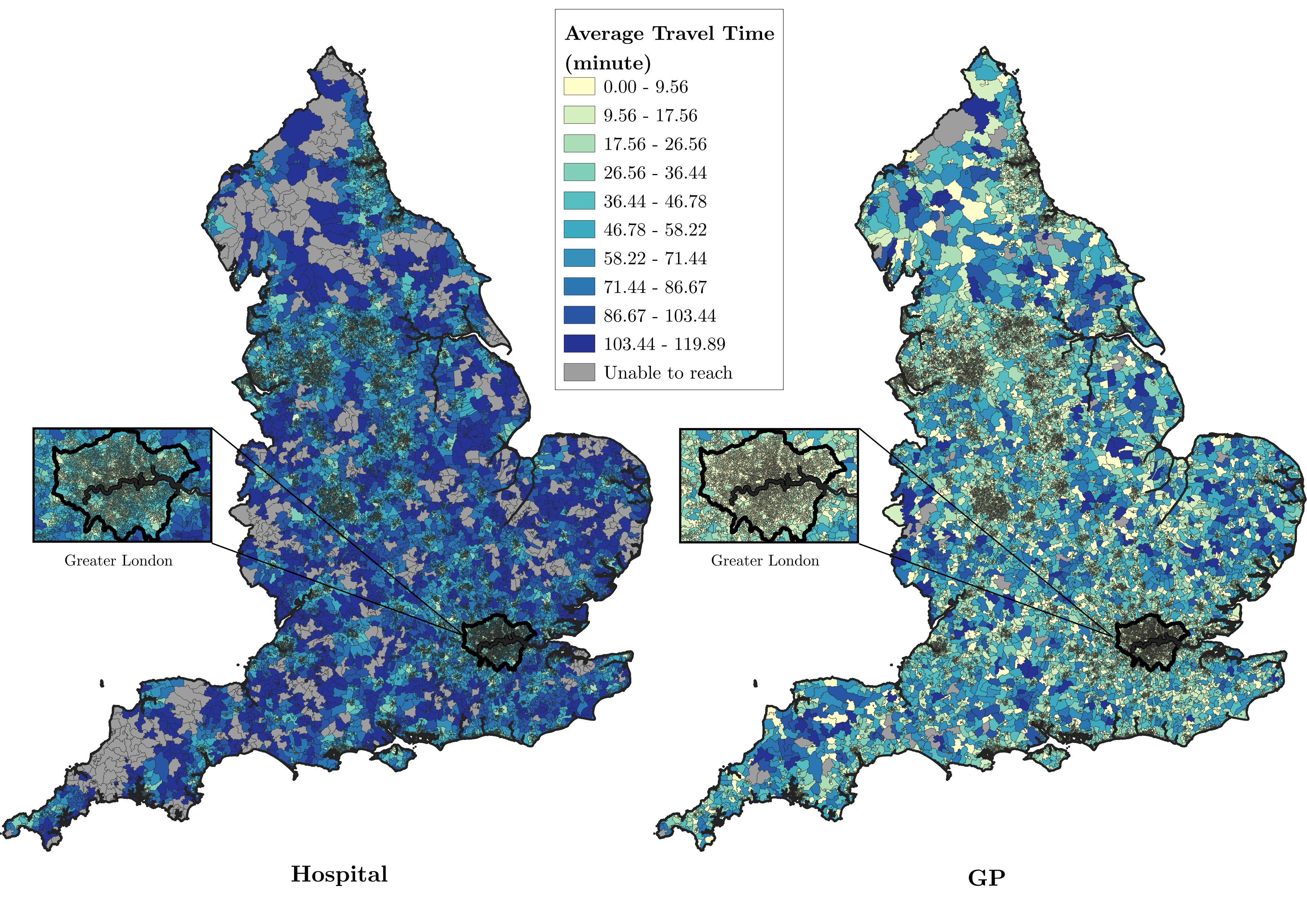}
        \caption{Average travel time from LSOAs in England to hospitals and GPs}
        \label{fig:hosp_gp_mean_aug}
    \end{subfigure}
    \hfill 
    \begin{subfigure}[b]{\textwidth}
        \centering
        \includegraphics[width=0.8\linewidth]{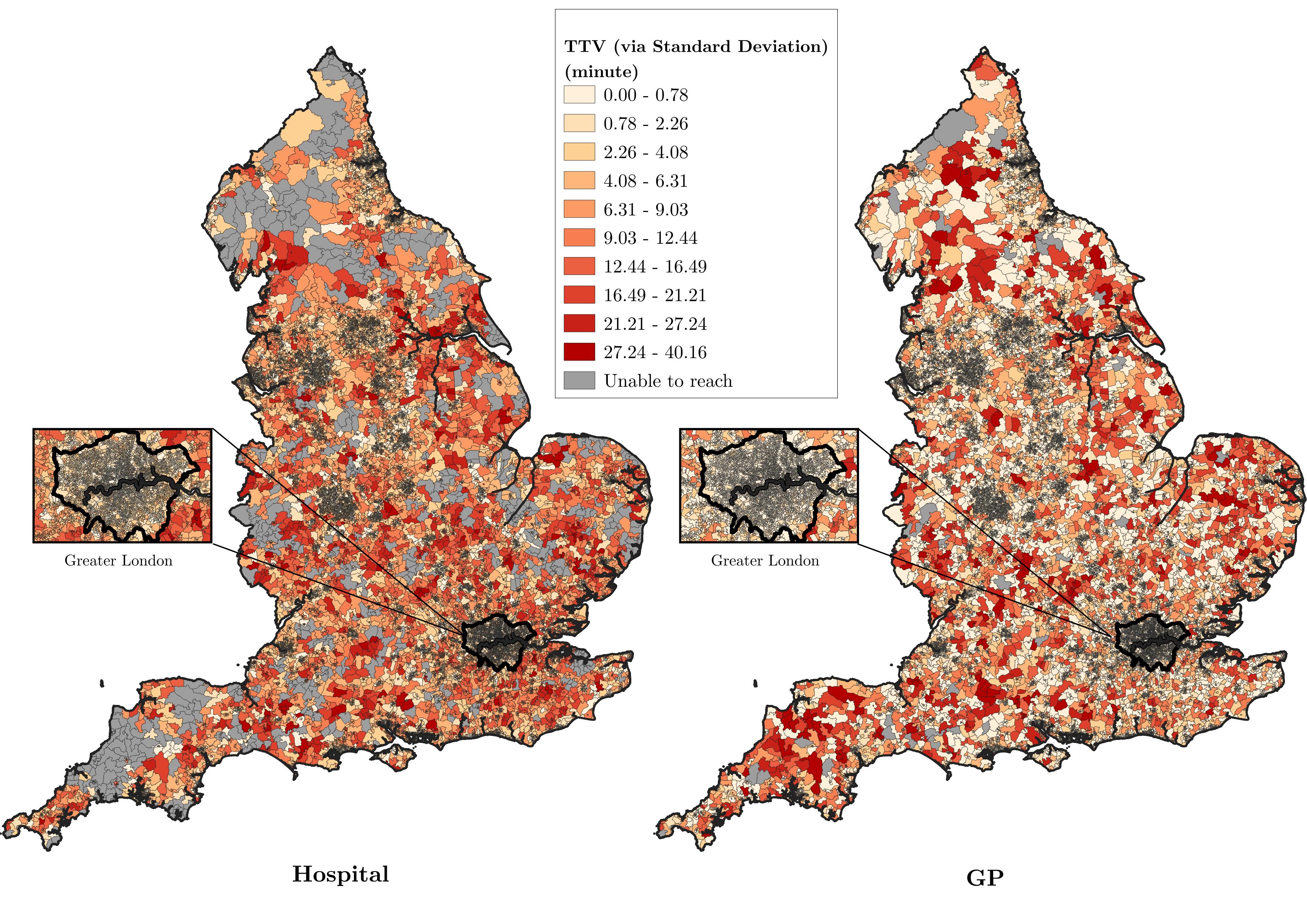}
        \caption{Travel time variability from LSOAs in England to hospitals and GPs}
        \label{fig:hosp_gp_sd_aug}
    \end{subfigure}
    \caption{Maps of LSOAs in England showing (a) average travel time and (b) travel time variability (TTV) to hospitals and GPs on 30th August.}
    \label{fig:ttv_aug}
\end{figure}

\begin{figure}[H]
    \centering
    \begin{subfigure}[b]{\textwidth}
        \centering
        \includegraphics[width=0.8\linewidth]{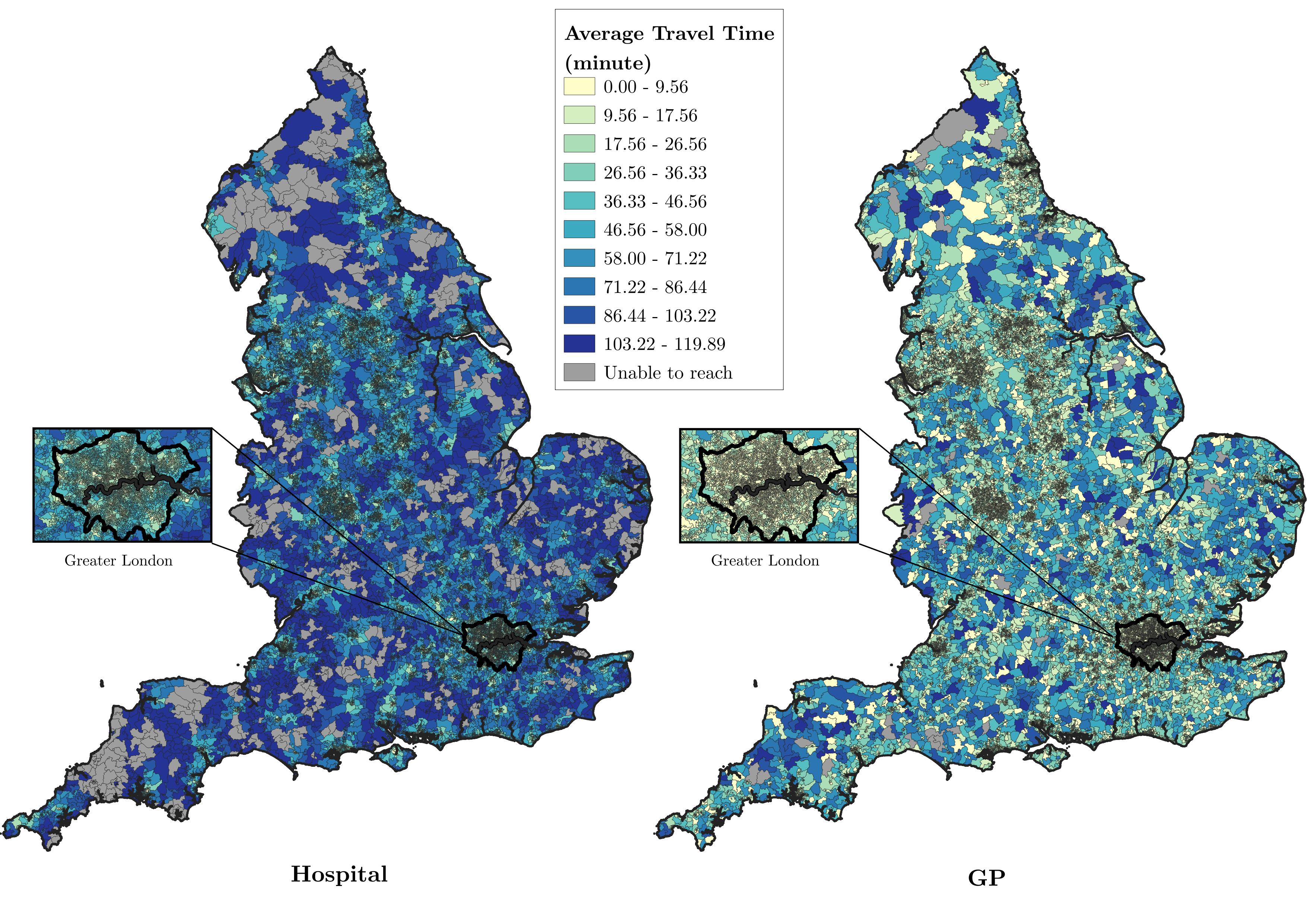}
        \caption{Average travel time from LSOAs in England to hospitals and GPs}
        \label{fig:hosp_gp_mean_nov}
    \end{subfigure}
    \hfill 
    \begin{subfigure}[b]{\textwidth}
        \centering
        \includegraphics[width=0.8\linewidth]{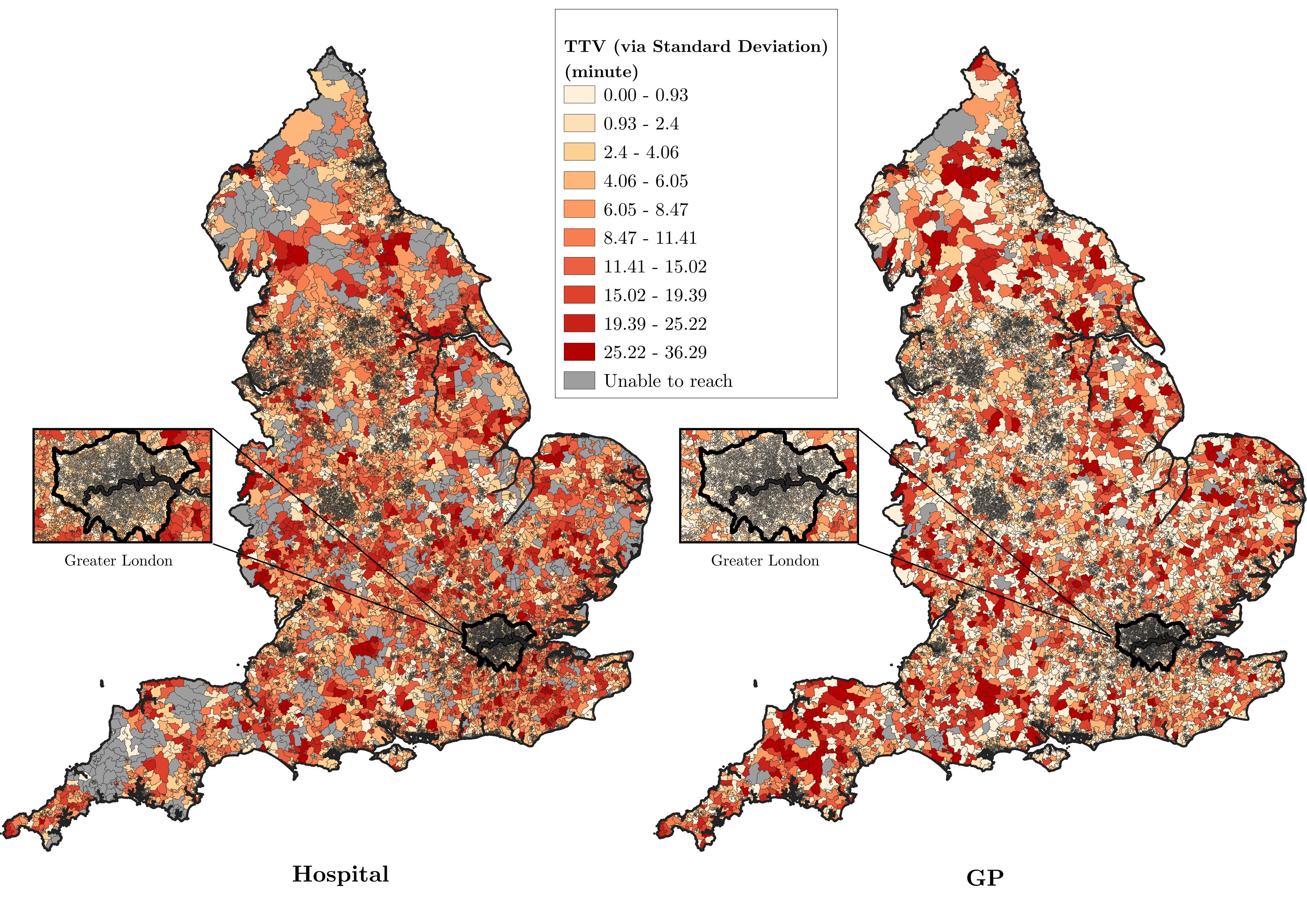}
        \caption{Travel time variability from LSOAs in England to hospitals and GPs}
        \label{fig:hosp_gp_sd_nov}
    \end{subfigure}
    \caption{Maps of LSOAs in England showing (a) average travel time and (b) travel time variability (TTV) to hospitals and GPs on 29th November.}
    \label{fig:ttv_nov}
\end{figure}

\end{document}